\documentclass{article}
\usepackage{smc2019}
\usepackage{times}
\usepackage{ifpdf}
\usepackage[english]{babel}
\usepackage{cite}

\usepackage{amsmath,url}
\usepackage{color}
\usepackage{enumitem}
\usepackage{amssymb,amsfonts}
\usepackage{algorithmic}
\usepackage{textcomp}

\usepackage{nicefrac}
\usepackage{pgfplots}
\usepackage{booktabs,siunitx}
\usepackage{subfig}
\pgfplotsset{compat=1.9}
\usepgfplotslibrary{units}
\usetikzlibrary{patterns}

\def\shallow{{\texttt{Shallow}}}
\def\deep{{\texttt{Deep}}}
\def\classmod{{\texttt{ClassMod}}}
\def\deepmod{{\texttt{DeepMod}}}
\def\shallowmod{{\texttt{ShallowMod}}}

\def\deepkey{{\texttt{DeepSpec}}}
\def\deeptempo{{\texttt{DeepTemp}}}
\def\deepsquare{{\texttt{DeepSquare}}}

\def\shallowkey{{\texttt{ShallowSpec}}}
\def\shallowtempo{{\texttt{ShallowTemp}}}
\def\halfshallowkey{{\texttt{HalfShallowSpec}}}
\def\halfshallowtempo{{\texttt{HalfShallowTemp}}}

\def\accone{{\emph{Accuracy1}}}
\def\acctwo{{\emph{Accuracy2}}}

\def\ballroom{{\texttt{Ballroom}}}
\def\eball{{\texttt{EBall}}}

\def\gtzan{{\texttt{GTzan}}}
\def\gtzankey{{\texttt{GTzan\,Key}}}
\def\gtzantempo{{\texttt{GTzan\,Tempo}}}

\def\giantsteps{{\texttt{GiantSteps}}}
\def\giantstepstempo{{\texttt{GiantSteps\,Tempo}}}
\def\giantstepskey{{\texttt{GiantSteps\,Key}}}

\def\lmd{{\texttt{LMD}}}
\def\lmdtempo{{\texttt{LMD\,Tempo}}}
\def\lmdkey{{\texttt{LMD\,Key}}}

\def\mtgkey{{\texttt{MTG\,Key}}}
\def\mtgtempo{{\texttt{MTG\,Tempo}}}

\def\papertitle{Musical Tempo and Key Estimation using Convolutional Neural Networks with Directional Filters}
\def\firstauthor{Hendrik Schreiber}
\def\secondauthor{Meinard M{\"u}ller}

\newif\ifpdf
\ifx\pdfoutput\relax
\else
   \ifcase\pdfoutput
      \pdffalse
   \else
      \pdftrue
\fi

\ifpdf 
  \usepackage[pdftex,
    pdftitle={\papertitle},
    pdfauthor={\firstauthor, \secondauthor},
    bookmarksnumbered, 
    pdfstartview=XYZ 
   ]{hyperref}

  \usepackage[figure,table]{hypcap}

\else 
  \usepackage[dvips,
    bookmarksnumbered, 
    pdfstartview=XYZ 
  ]{hyperref}  

  \usepackage[dvips]{epsfig,graphicx}
  \graphicspath{{./figures/}}
  \DeclareGraphicsExtensions{.eps}

  \usepackage[figure,table]{hypcap}
\fi

\hypersetup{
    colorlinks,%
    citecolor=black,%
    filecolor=black,%
    linkcolor=black,%
    urlcolor=black
}

\title{\papertitle}

 \twoauthors
   {\firstauthor} {tagtraum industries incorporated \\ %
     {\tt \href{mailto:hs@tagtraum.com}{hs@tagtraum.com}}}
   {\secondauthor} {International Audio Laboratories Erlangen \\ %
     {\tt \href{mailto:meinard.mueller@audiolabs-erlangen.de}{meinard.mueller@audiolabs-erlangen.de}}}

\begin{document}
\capstartfalse
\maketitle
\capstarttrue

\begin{abstract}
In this article we explore how the different semantics of spectrograms' time and
frequency axes can be exploited for musical tempo and key estimation
using Convolutional Neural Networks (CNN).
By addressing both tasks with the same network architectures ranging from shallow,
domain-specific approaches to deep variants with directional filters, 
we show that axis-aligned architectures perform similarly well as
common VGG-style networks developed for computer vision, while being less vulnerable to
confounding factors and requiring fewer model parameters.
\end{abstract}

\section{Introduction}
\label{sec:introduction}

\begin{figure}[t]
 \centerline{
 \includegraphics[width=7cm]{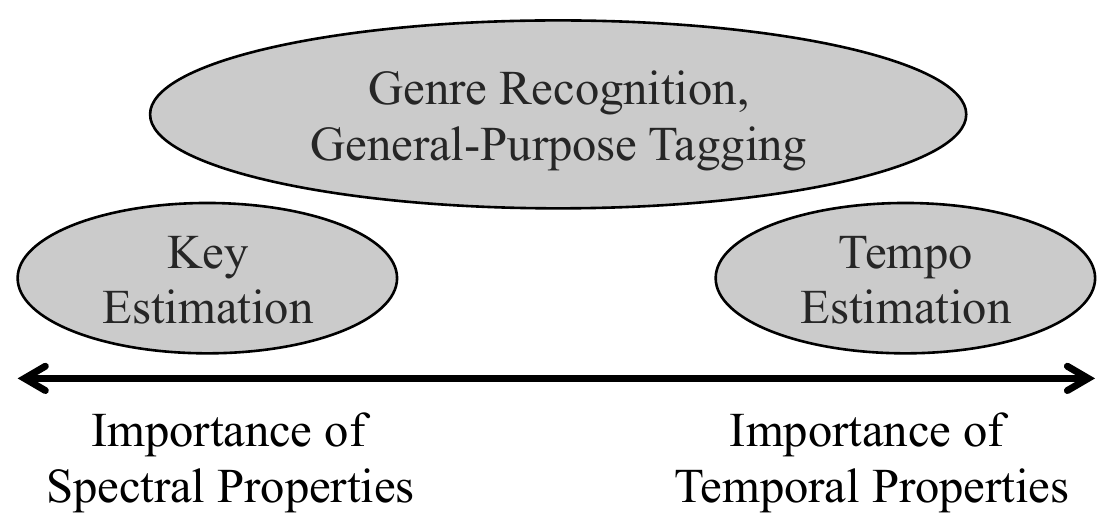}}
 \caption{Several MIR tasks and their reliance on spectral or temporal signal properties.}
 \label{fig:mir-tasks}
\end{figure}

In recent years Convolutional Neural Networks~(CNN) have been employed for various
Music Information Retrieval~(MIR) tasks, such as
key detection~\cite{Korzeniowski2017,Korzeniowski2018},
tempo estimation~\cite{Schreiber2018a},
beat and rhythm analysis~\cite{holzapfel2016bayesian,durand2017robust,Gkiokas2017},
genre recognition~\cite{Dieleman2011,Schindler2016}, and
general-purpose tagging~\cite{Dorfer2018,Choi2016}.
Typically, a spectrogram is fed to the CNN and then classified in a way
appropriate for the task.
In contrast to recent computer vision approaches like Oxford's Visual Geometry Group's~(VGG)
deep image recognition network~\cite{Simonyan2014}, some of the employed CNN architectures
for MIR tasks use rectangular instead of square filters. The underlying idea is that, while
for  images the axes \emph{width} and \emph{height} have the same meaning, the spectrogram
axes \emph{frequency} and \emph{time} have fundamentally different meaning.
For MIR tasks mainly concerned with temporal aspects of
music~(e.g., tempo estimation, rhythmic patterns), rectangular filters aligned
with the time axis
appear suitable~\cite{Schreiber2018a}. Correspondingly,
tasks primarily concerned with frequency content~(e.g., chord or key detection),
may be approached with rectangular filters aligned with the frequency axis~\cite{McFee2017}.
In fact, tempo and key estimation can be seen as tasks from two different ends of a spectrum
of common MIR tasks, which are addressed by systems relying more or less on temporal or
spectral signal properties~(\figref{fig:mir-tasks}). Systems for other
tasks like general-purpose tagging or genre recognition are found more towards the
center of this spectrum as they usually require both spectral and temporal information.

In~\cite{Pons2016} Pons et al.\ explored the role of CNN filter shapes for MIR tasks.
In particular, they examined using rectangular filters in shallow CNNs for automatic
genre recognition of ballroom tracks.
Defining \emph{temporal} filter shapes as $1\times n$ and \emph{spectral} filter shapes
as $m\times 1$, they showed that using temporal filters alone, $81.8\%$ accuracy can
be reached, which is in line with a Nearest Neighbour classifier (k-NN) using
tempo as feature scoring $82.3\%$~\cite{GouyonDPW04_GenreRhythm_AES}.
Using just spectral filters, the test network reached $59.6\%$ accuracy, and a fusion
architecture with both temporal and spectral filters performed as well as an architecture
using square filters, scoring $87\%$.
The experiments confirmed that such \emph{directional} filters can be used to match either
temporal or spectral signal properties and that both may be useful for genre recognition.

Even though directional filters did not outperform square filters, there
are good arguments for using them:
First, CNNs using specialized, directional filters may use fewer
parameters or match musical concepts using fewer layers~\cite{Pons2017}.
Second, by limiting what a filter can match, one can influence what a CNN might learn,
thus better avoid ``horses''~\cite{Sturm2014simple} and improve explainability.
The latter is especially interesting for genre recognition systems, given their 
somewhat troubled history with respect to explicit matching of
musical concepts~\cite{Sturm2013,Pons2017}.
To further explore how and why directional or square filters contribute to results
achieved by CNN-based classification systems for MIR tasks, we believe it is beneficial to
build on Pons et al.'s work and experiment with tasks that explicitly aim to recognize either
high-level temporal or spectral properties, avoiding hard to define concepts like genre.
Such tasks are global key and tempo estimation.

%
%

The remainder of this paper is structured as follows:
In \secref{sec:experiments} we describe our experiments by defining both tasks,
the used network variants, the training procedure, and evaluation.
The results are then presented in \secref{sec:results} and discussed in
\secref{sec:discussion}.
Finally, in \secref{sec:conclusions} we present our conclusions.

\section{Experiments}
\label{sec:experiments}

For the purpose of comparing the effects of using different filter shapes we train and
evaluate different CNN architectures for the key and tempo estimation tasks using
several datasets. In this section, we first describe the two tasks, then discuss the
used network architectures and datasets, and finally outline the evaluation procedure.

\subsection{Key Estimation}
\label{sec:keyestimation}

Key estimation attempts to predict the correct key for a given piece of music.
Oftentimes, the problem is restricted to major and minor modes,
ignoring other possible modes like Dorian or Lydian,
and to pieces without modulation. Framed this way, key estimation is a
classification problem with $N_\mathrm{K}=24$ different classes (12 tonics, major/minor).
The current state-of-the-art system is CNN-based using a VGG-style
architecture with square filters~\cite{Korzeniowski2018} and a fully
convolutional classification stage, as opposed to a fully connected one.
This allows training on short and prediction on variable length spectrograms. 

In our experiments we follow a similar approach. As input to the
network~(\secref{sec:architectures}) we use constant-Q magnitude spectrograms
with the dimensions $F_\mathrm{K} \times T_\mathrm{K} = 168 \times 60$; $F_\mathrm{K}$ being the
number of frequency bins and $T_\mathrm{K}$ the number of time frames. $F_\mathrm{K}$ covers the
frequency range of $7$ octaves with a frequency resolution of two bins per semitone.
Time resolution is $0.19\,\mathrm{s}$ per time frame, i.e.\ $60$ frames
correspond to $11.1\,\mathrm{s}$.
Since all training samples are longer than $11.1\,\mathrm{s}$, we choose a random offset
for each sample during each training epoch and crop the spectrogram to $60$ frames.
To account for class imbalances within the two modes, each spectrogram is randomly shifted 
along the frequency axis by $\{-4, -3, \ldots, 6, 7\}$ semitones and the ground truth
labels are adjusted accordingly. We define \emph{no shift} to correspond to a
spectrogram covering the $7$ octaves starting at pitch E1.
In practice, we simply crop an $8$ octaves spanning spectrogram that starts at C1 to
$7$ octaves. After cropping the spectrogram is normalized so that it has zero mean
and unit variance.

\subsection{Tempo Estimation}
\label{sec:tempoestimation}

Even though tempo estimation naturally appears to be a regression task, Schreiber and
M{\"u}ller~\cite{Schreiber2018a} have shown that it can also be treated as a classification
task by mapping Beats Per Minute~(BPM) values to distinct tempo classes.
Concretely, their system maps the integer tempo values $\{30,\ldots,285\}$ to $N_\mathrm{T}=256$
classes. As input to a CNN with temporal filters and elements from~\cite{Szegedy2015going}
and~\cite{Pons2017} they use mel-magnitude-spectrograms. 
Even though we work with other network architectures
than~\cite{Schreiber2018a}~(\secref{sec:architectures}),
we use the same general setup. We also treat tempo estimation
as classification into $256$ classes and use mel-magnitude-spectrograms with the
dimensions $F_\mathrm{T} \times T_\mathrm{T} = 40 \times 256$ as input;
$F_\mathrm{T}$ being the number of frequency bins and $T_\mathrm{T}$ the number of time frames.
$F_\mathrm{T}$ covers the frequency range $20 - 5,000\,\mathrm{Hz}$. The time resolution is
$0.46\,\mathrm{ms}$ per time frame, i.e., $256$ frames correspond to $11.9\,\mathrm{s}$. 

Just like the training excerpts for key estimation, the mel-spectrograms are cropped
to the right size using a different randomly chosen offset during each epoch.
To augment the training dataset, spectrograms are scaled along the time axis before
cropping using the factors $\{0.8, 0.84, \ldots, 1.16, 1.2\}$. Ground truth labels
are adjusted accordingly~\cite{Schreiber2018a}.
After cropping and scaling spectrograms are normalized ensuring
zero mean and unit variance per sample.

\begin{table}[t]
\centering
\subfloat[\shallow]{
\begin{tabular}{l} \toprule
Module \\ \midrule
\shallowmod\ 
\\
\\
\\
\\
\\ \\\midrule

\classmod\ \\
\bottomrule

\end{tabular}
\label{tab:architectures-a}
}
\quad
\subfloat[\deep]{
\begin{tabular}{lc} \toprule
Module         & Size \\ \midrule
\deepmod\        & $\ell=0$  \\
\deepmod\        & $\ell=1$  \\
\deepmod\        & $\ell=2$  \\
\deepmod\        & $\ell=2$  \\
\deepmod\        & $\ell=3$  \\
\deepmod\        & $\ell=3$  \\ \midrule
\classmod\ \\
\bottomrule
\end{tabular}
\label{tab:architectures-b}
}
\caption{Used network architectures. (a) \shallow\ architecture consisting of
a variant of the \shallowmod\
module and a \classmod\ module. (b) \deep\ architecture consisting of multiple,
\deepmod\ modules parameterized with $\ell$ to influence the filter count
and a \classmod\ module.}
\label{tab:architectures}
\end{table}

\begin{table}[t]
\small{
\centering
\begin{tabular}{lccc} 
\multicolumn{2}{l}{(a) \shallowmod} \\ \toprule
Layer         & Temp   & Spec & Square      \\ \midrule
Input \\
Conv, ReLU    & $k$, $1 \times 3$  & $k$, $3 \times 1$  & n.a.  \\
Dropout       & $p_\mathrm{D}$ & $p_\mathrm{D}$ & n.a. \\
AvgPool       & $F_\mathrm{T} \times 1$ & $1 \times T_\mathrm{K}$  & n.a.\\
Conv, ReLU    & $64k$, $1 \times T_\mathrm{T}$ & $64k$, $F_\mathrm{K} \times 1$  & n.a. \\
Dropout       & $p_\mathrm{D}$ & $p_\mathrm{D}$ & n.a.             \\ \bottomrule
\\
\multicolumn{2}{l}{(b) \deepmod} \\ \toprule
Layer         & Temp   & Spec & Square      \\ \midrule
Input \\
Conv, ReLU    & $2^\ell k$, $1 \times 5$ & $2^\ell k$, $5 \times 1$ & $2^\ell k$, $5 \times 5$ \\
BatchNorm     & \\
Conv, ReLU    & $2^\ell k$, $1 \times 3$ & $2^\ell k$, $3 \times 1$ & $2^\ell k$, $3 \times 3$ \\
BatchNorm     & \\
MaxPool       & $2 \times 2$            & $2 \times 2$            & $2 \times 2$ \\
Dropout       & $p_\mathrm{D}$ & $p_\mathrm{D}$ & $p_\mathrm{D}$       \\ \bottomrule
\\
\multicolumn{2}{l}{(c) \classmod} \\ \toprule
Layer         & Temp   & Spec & Square      \\ \midrule
Input \\
Conv, ReLU    & $N_\mathrm{T}$, $1\times1$ & $N_\mathrm{K}$, $1\times1$ & n.a. \\
GlobalAvgPool &             &             \\
Softmax       &             &             \\ \bottomrule
\end{tabular}
\caption{Layer definitions for the three modules \shallowmod, \classmod, and \deepmod,
describing number of filters (e.g., $k$ or $64k$)
and their respective shapes (e.g., $1 \times 3$ or $5 \times 5$).}
\label{tab:modules}
}
\end{table}

\subsection{Network Architectures}
\label{sec:architectures}

To gain insights into how filter shapes affect performance of CNN-based key and tempo
estimation systems we run experiments with two very different architectures:
a relatively shallow but specialized one, and a commonly used much deeper one from
the field of computer vision. Both architectures are used for both tasks.

\subsubsection{Shallow Architectures}

Our \shallow\ architectures, outlined in \tabref{tab:architectures-a},
consists of two parts: the feature extraction module \shallowmod\
and the classification module \classmod.
\shallowmod, depicted in~\tabref{tab:modules}a, is inspired by a classic signal processing approach
that first attempts to find local spectrogram peaks along one axis, averages these
peaks over the other axis, and then attempts to find a global pattern, i.e., a periodicity
for tempo estimation~\cite{Klapuri99_OnsetDetection_ICASSP} and a
pitch profile for key detection~\cite{Krumhansl90}.
In terms of CNNs this means that our first convolutional layer consists of short
directional filters (local peaks), followed by a one-dimensional average pooling layer
that is orthogonal to the short filters, followed by a layer
with long directional filters~(global pattern) that stretch in the same direction
as the short filters.
We use $\mathrm{ReLU}$ as activation function for the convolutional
layers and to avoid overfitting we add a dropout layer~\cite{Srivastava2014dropout} after
each $\mathrm{ReLU}$.
The parameters $k$ and $p_\mathrm{D}$ let us scale the number of convolutional filters 
and dropout probabilities.
\shallowmod\ is followed by a fully convolutional
classification module named \classmod~(\tabref{tab:modules}c), which consists of
a $1\times1$ bottleneck layer (pointwise convolution) with as many filters as desired
classes ($N_\mathrm{K}$ or $N_\mathrm{T}$), a global average pooling layer, and the
$\mathrm{softmax}$ activation function. 
Note, that because all directional filters are identically aligned, the model has
an asymmetric, \emph{directional capacity}, i.e., it has a much larger ability to
describe complex relationships in one direction than in the other.

We use the same general architecture for both key and tempo estimation.
The only differences are the filter and pooling directions and dimensions.
For tempo estimation we use temporal filters with pooling along the
frequency axis, and for key estimation spectral filters with
pooling along the time axis.
Both architectures are named after their filter directions, \shallowtempo\
and \shallowkey, respectively.
We also adjust the pooling and the long filters shape to the size of the input
spectrogram, which is different for the two tasks.

\subsubsection{Deep Architectures}

The second architecture, \deep~(\tabref{tab:architectures-b}), is a common
VGG-style architecture consisting of six parameterized
feature extraction modules \deepmod~(\tabref{tab:modules}b) followed by the same
classification module that we have already used in \shallow.
Each of the feature extraction modules contains a convolutional layer with $5\times5$
filters followed by a convolutional layer with $3\times3$ filters.
The convolutional layers consist of $2^\ell k$ filters each, with network parameter $k$
and module parameter $\ell$.
While $\ell$ influences the number of filters in an instance of \deepmod,
$k$ lets us scale the total number of parameters in the network. As shown in
\tabref{tab:architectures-b}, deeper instances have more filters.
The convolutional layers are followed by a $2\times2$ max pooling layer. Should
pooling not be possible along an axis, because
the output from the previous layer is only $1$ wide or high, pooling is skipped
along that axis. This happens for example, when a tempo spectrogram with its $40$ bands
passes through more than $5$ max pools.
Each pooling layer is followed by a dropout layer with probability $p_\mathrm{D}$.
To counter covariate shift, we add batch normalization~\cite{Ioffe2015} layers
after each convolutional layer.

The general structure of the \deep\ architecture is customized neither for the key nor
for the tempo task. However, in order to investigate how different filter shapes affect
the network's performance, we modify the described architecture by replacing the
square convolutional kernels with directional ones, i.e., $3\times3$ with $1\times3$ or
$3\times1$, and $5\times5$ with $1\times5$ or $5\times1$.
Analogous to the naming scheme used for shallow networks, 
we denote the directional variants \deeptempo\ and \deepkey.
The original variant is named \deepsquare.

\subsection{Datasets}

\begin{table}[t]
\centering
\begin{tabular}{rl} \toprule
Split      & Key Datasets\\ \midrule
Training   & $80\%$ of \lmdkey\ $\cup$ $80\%$ of \mtgkey\ \\
Validation & $10\%$ of \lmdkey\ $\cup$ $20\%$ of \mtgkey\ \\
Testing    & \giantstepskey, \gtzankey, \\
           & $10\%$ of \lmdkey\ \\
\bottomrule
\\
\toprule
Split      & Tempo Datasets\\ \midrule
Training   & $80\%$ of \eball\ $\cup$ $80\%$ of \mtgtempo\ \\
           & $\cup$ $80\%$ of \lmdtempo\ \\
Validation & $20\%$ of \eball\ $\cup$ $20\%$ of \mtgtempo\ \\
           & $\cup$ $10\%$ of \lmdtempo\ \\
Testing    & \giantstepstempo, \gtzantempo, \\
           & $10\%$ of \lmdtempo, \ballroom\ \\
\bottomrule
\end{tabular}
\caption{Dataset splits used in key (top) and tempo (bottom) estimation experiments.}
\label{tab:splits}
\end{table}

We use the following publicly available datasets from different genres
for both training and evaluation (listed in alphabetical order).
The used splits are randomly chosen and listed in \tabref{tab:splits}.

\begin{description}[leftmargin=0.5cm,font=\normalfont]

\item[\ballroom\,($698$\,samples):] $30\,\mathrm{s}$ excerpts with tempo annotations~\cite{GouyonKDATUC06_experimentalTempoComparison_TASL}.

\item[\eball\,($3{,}826$\,samples):] $30\,\mathrm{s}$ excerpts with tempo annotations,
excluding tracks also occurring in the regular \ballroom\ dataset~\cite{Marchand2016,GouyonKDATUC06_experimentalTempoComparison_TASL,Schreiber2018a}.

\item[\giantstepskey\ ($604$\,samples):] $2\,\mathrm{min}$ excerpts of electronic
dance music~(EDM)~\cite{knees_etal:ismir:2015}.

\item[\giantstepstempo\,($661$\,samples):] $2\,\mathrm{min}$ excerpts of
EDM~\cite{knees_etal:ismir:2015}.
Revised tempo annotations from~\cite{Schreiber2018b}.

\item[\gtzankey\,($836$\,samples):] $30\,\mathrm{s}$ excerpts from $10$ different genres~\cite{TzanetakisC02_GenreRecognition_TASL}.
Key annotations from~\cite{Kraft2013tonalness}.\footnote{\url{https://github.com/alexanderlerch/gtzan_key}} Most tracks with missing key annotations belong to the
genres classical, jazz, and hip-hop.

\item[\gtzantempo\,($999$\,samples):] $30\,\mathrm{s}$ excerpts from $10$ different genres~\cite{TzanetakisC02_GenreRecognition_TASL}.
Tempo annotations from~\cite{percival2014streamlined}.

\item[\lmdkey\ ($6{,}981$\,samples):] $30\,\mathrm{s}$ excerpts, predominantly rock and pop~\cite{Raffel2016,Schreiber2017key}.
Due to a MIDI peculiarity, this dataset does not contain any tracks in C\,major. 
Some form of data augmentation as described above is therefore necessary.

\item[\lmdtempo\,($3{,}611$\,samples):] $30\,\mathrm{s}$ excerpts, predominantly rock and pop~\cite{Raffel2016,Schreiber2018a}.

\item[\mtgtempo\ / \mtgkey\ ($1{,}158$\,samples):] $2\,\mathrm{min}$ EDM excerpts
annotated with both key and
tempo~\cite{Faraldo2017multi,Schreiber2018a}. We used only tracks that are still
publicly available, have an unambiguous key, and a high key annotation
confidence.\footnote{\url{https://github.com/GiantSteps/giantsteps-mtg-key-dataset}} 

\end{description}

\subsection{Evaluation}

Since the proposed network architectures are fully convolutional, we can
choose at prediction time to pass a track either in one long spectrogram
or as multiple shorter windows through the network.
In the latter case, predictions for all windows would have to
be aggregated. Informal experiments did not show a remarkable difference. For this
work we choose to predict values for whole spectrograms.

When evaluating key estimation systems either a simple accuracy or a score is used
that assigns additional value to musically justifiable mistakes, like being off by a
perfect fifth.\footnote{\url{https://www.music-ir.org/mirex/wiki/2018:Audio_Key_Detection}}
For this work, we choose to only report the percentage of correctly classified keys.
Tempo estimation systems are typically evaluated using the metrics \accone\ and \acctwo.
While \accone\ reports the percentage of correctly estimated tempi allowing a $4\%$
tolerance, \acctwo\ additionally permits so-called octave errors, i.e., errors by
a factor of $2$ and $3$~\cite{GouyonKDATUC06_experimentalTempoComparison_TASL}.
We choose to report only \accone.

For training we use Adam~\cite{Kingma2014} as optimizer with a learning rate of
$0.001$, a batch size of $32$, and early stopping once the validation loss has not
decreased any more during the last $100$ epochs. In this work, one epoch is defined as
having shown all training samples to the network once, regardless of augmentation.
We choose $k$ so that we can compare architectures with similar parameter counts.
\shallow\ is trained with $k \in \{1, 2, 4, 6, 8, 12\}$
and \deep\ with $k \in \{2, 4, 8, 16, 24\}$. Additionally, \deepsquare\ is
trained with $k=1$.
For both architectures we apply various dropout probabilities
$p_\mathrm{D} \in \{0.1, 0.3, 0.5\}$. 
Each variant is trained $5$ times and mean validation accuracy along with its
standard deviation is recorded for each variant. In total we train $420$ models
with $84$ different configurations.

For testing, we pick the dropout variant of each network class that performed best on 
the validation set and evaluate it against the test datasets. Again, we report the mean
accuracies for $5$ runs along with their standard deviations.

\section{Results}
\label{sec:results}


\pgfplotstableread[row sep=\\,col sep=&]{
Testset & Runs & Parameters & Acc1 & Std1 & Model \\
tempo_valid      & 5 &      19268 & 0.10052 & 0.02886 & shallow_key_in=(40__256__1)_out=256_filters=1_short_shape=(3__1)_long_shape=(40__1)_dropout=0.1 \\
tempo_valid      & 5 &      43400 & 0.11363 & 0.05759 & shallow_key_in=(40__256__1)_out=256_filters=2_short_shape=(3__1)_long_shape=(40__1)_dropout=0.3 \\
tempo_valid      & 5 &     107024 & 0.11982 & 0.01944 & shallow_key_in=(40__256__1)_out=256_filters=4_short_shape=(3__1)_long_shape=(40__1)_dropout=0.3 \\
tempo_valid      & 5 &     191128 & 0.12704 & 0.01560 & shallow_key_in=(40__256__1)_out=256_filters=6_short_shape=(3__1)_long_shape=(40__1)_dropout=0.3 \\
tempo_valid      & 5 &     295712 & 0.12955 & 0.01340 & shallow_key_in=(40__256__1)_out=256_filters=8_short_shape=(3__1)_long_shape=(40__1)_dropout=0.1 \\
tempo_valid      & 5 &     566320 & 0.13029 & 0.03261 & shallow_key_in=(40__256__1)_out=256_filters=12_short_shape=(3__1)_long_shape=(40__1)_dropout=0.5 \\
}\tempokeyshallow

\pgfplotstableread[row sep=\\,col sep=&]{
Testset & Runs & Parameters & Acc1 & Std1 & Model \\
tempo_valid      & 5 &      17988 & 0.08298 & 0.02174 & shallow_key_in=(40__256__1)_out=256_filters=1_short_shape=(3__1)_long_shape=(20__1)_dropout=0.3 \\
tempo_valid      & 5 &      38280 & 0.12970 & 0.01401 & shallow_key_in=(40__256__1)_out=256_filters=2_short_shape=(3__1)_long_shape=(20__1)_dropout=0.3 \\
tempo_valid      & 5 &      86544 & 0.13692 & 0.04566 & shallow_key_in=(40__256__1)_out=256_filters=4_short_shape=(3__1)_long_shape=(20__1)_dropout=0.3 \\
tempo_valid      & 5 &     145048 & 0.11614 & 0.02428 & shallow_key_in=(40__256__1)_out=256_filters=6_short_shape=(3__1)_long_shape=(20__1)_dropout=0.1 \\
tempo_valid      & 5 &     213792 & 0.13265 & 0.02105 & shallow_key_in=(40__256__1)_out=256_filters=8_short_shape=(3__1)_long_shape=(20__1)_dropout=0.5 \\
tempo_valid      & 5 &     382000 & 0.13721 & 0.04780 & shallow_key_in=(40__256__1)_out=256_filters=12_short_shape=(3__1)_long_shape=(20__1)_dropout=0.5 \\
}\tempokeyshallowhalf

\pgfplotstableread[row sep=\\,col sep=&]{
Testset & Runs & Parameters & Acc1 & Std1 & Model \\
tempo_valid      & 5 &      33092 & 0.69270 & 0.31820 & shallow_tempo_in=(40__256__1)_out=256_filters=1_short_shape=(1__3)_long_shape=(1__256)_dropout=0.1 \\
tempo_valid      & 5 &      98696 & 0.88902 & 0.01549 & shallow_tempo_in=(40__256__1)_out=256_filters=2_short_shape=(1__3)_long_shape=(1__256)_dropout=0.3 \\
tempo_valid      & 5 &     328208 & 0.90833 & 0.00815 & shallow_tempo_in=(40__256__1)_out=256_filters=4_short_shape=(1__3)_long_shape=(1__256)_dropout=0.5 \\
tempo_valid      & 5 &     688792 & 0.91363 & 0.00614 & shallow_tempo_in=(40__256__1)_out=256_filters=6_short_shape=(1__3)_long_shape=(1__256)_dropout=0.5 \\
tempo_valid      & 5 &    1180448 & 0.91157 & 0.00707 & shallow_tempo_in=(40__256__1)_out=256_filters=8_short_shape=(1__3)_long_shape=(1__256)_dropout=0.3 \\
tempo_valid      & 5 &    2556976 & 0.91805 & 0.00157 & shallow_tempo_in=(40__256__1)_out=256_filters=12_short_shape=(1__3)_long_shape=(1__256)_dropout=0.5 \\
}\tempotemposhallow

\pgfplotstableread[row sep=\\,col sep=&]{
Testset & Runs & Parameters & Acc1 & Std1 & Model \\
tempo_valid      & 5 &      24900 & 0.70479 & 0.32193 & shallow_tempo_in=(40__256__1)_out=256_filters=1_short_shape=(1__3)_long_shape=(1__128)_dropout=0.3 \\
tempo_valid      & 5 &      65928 & 0.90052 & 0.00957 & shallow_tempo_in=(40__256__1)_out=256_filters=2_short_shape=(1__3)_long_shape=(1__128)_dropout=0.5 \\
tempo_valid      & 5 &     197136 & 0.91349 & 0.00318 & shallow_tempo_in=(40__256__1)_out=256_filters=4_short_shape=(1__3)_long_shape=(1__128)_dropout=0.3 \\
tempo_valid      & 5 &     393880 & 0.91231 & 0.00876 & shallow_tempo_in=(40__256__1)_out=256_filters=6_short_shape=(1__3)_long_shape=(1__128)_dropout=0.5 \\
tempo_valid      & 5 &     656160 & 0.92012 & 0.00444 & shallow_tempo_in=(40__256__1)_out=256_filters=8_short_shape=(1__3)_long_shape=(1__128)_dropout=0.3 \\
tempo_valid      & 5 &    1377328 & 0.91127 & 0.01081 & shallow_tempo_in=(40__256__1)_out=256_filters=12_short_shape=(1__3)_long_shape=(1__128)_dropout=0.5 \\
}\tempotemposhallowhalf

\pgfplotstableread[row sep=\\,col sep=&]{
Testset & Runs & Parameters & Acc1 & Std1 & Model \\
tempo_valid      & 5 &       9322 & 0.80929 & 0.01042 & vgg_like_in=(40__256__1)_out=256_filters=2_pool_shape=(2__2)_max=True_filter_shapes=[(1__5)__(1__3)]_dropout=0.1 \\
tempo_valid      & 5 &      27228 & 0.88357 & 0.00696 & vgg_like_in=(40__256__1)_out=256_filters=4_pool_shape=(2__2)_max=True_filter_shapes=[(1__5)__(1__3)]_dropout=0.1 \\
tempo_valid      & 5 &      89560 & 0.88681 & 0.01246 & vgg_like_in=(40__256__1)_out=256_filters=8_pool_shape=(2__2)_max=True_filter_shapes=[(1__5)__(1__3)]_dropout=0.1 \\
tempo_valid      & 5 &     320304 & 0.91570 & 0.00982 & vgg_like_in=(40__256__1)_out=256_filters=16_pool_shape=(2__2)_max=True_filter_shapes=[(1__5)__(1__3)]_dropout=0.1 \\
tempo_valid      & 5 &     692488 & 0.91349 & 0.00764 & vgg_like_in=(40__256__1)_out=256_filters=24_pool_shape=(2__2)_max=True_filter_shapes=[(1__5)__(1__3)]_dropout=0.3 \\
}\tempotempovgg

\pgfplotstableread[row sep=\\,col sep=&]{
Testset & Runs & Parameters & Acc1 & Std1 & Model \\
tempo_valid      & 5 &       9322 & 0.26249 & 0.03834 & vgg_like_in=(40__256__1)_out=256_filters=2_pool_shape=(2__2)_max=True_filter_shapes=[(5__1)__(3__1)]_dropout=0.1 \\
tempo_valid      & 5 &      27228 & 0.37097 & 0.06060 & vgg_like_in=(40__256__1)_out=256_filters=4_pool_shape=(2__2)_max=True_filter_shapes=[(5__1)__(3__1)]_dropout=0.1 \\
tempo_valid      & 5 &      89560 & 0.53073 & 0.03545 & vgg_like_in=(40__256__1)_out=256_filters=8_pool_shape=(2__2)_max=True_filter_shapes=[(5__1)__(3__1)]_dropout=0.1 \\
tempo_valid      & 5 &     320304 & 0.56654 & 0.06851 & vgg_like_in=(40__256__1)_out=256_filters=16_pool_shape=(2__2)_max=True_filter_shapes=[(5__1)__(3__1)]_dropout=0.1 \\
tempo_valid      & 5 &     692488 & 0.63228 & 0.06235 & vgg_like_in=(40__256__1)_out=256_filters=24_pool_shape=(2__2)_max=True_filter_shapes=[(5__1)__(3__1)]_dropout=0.1 \\
}\tempokeyvgg

\pgfplotstableread[row sep=\\,col sep=&]{
Testset & Runs & Parameters & Acc1 & Std1 & Model \\
tempo_valid      & 5 &       7134 & 0.77185 & 0.00366 & vgg_like_in=(40__256__1)_out=256_filters=1_pool_shape=(2__2)_max=True_filter_shapes=[(5__5)__(3__3)]_dropout=0.1 \\
tempo_valid      & 5 &      23082 & 0.86971 & 0.01106 & vgg_like_in=(40__256__1)_out=256_filters=2_pool_shape=(2__2)_max=True_filter_shapes=[(5__5)__(3__3)]_dropout=0.1 \\
tempo_valid      & 5 &      82188 & 0.91688 & 0.00428 & vgg_like_in=(40__256__1)_out=256_filters=4_pool_shape=(2__2)_max=True_filter_shapes=[(5__5)__(3__3)]_dropout=0.1 \\
tempo_valid      & 5 &     309240 & 0.92822 & 0.00568 & vgg_like_in=(40__256__1)_out=256_filters=8_pool_shape=(2__2)_max=True_filter_shapes=[(5__5)__(3__3)]_dropout=0.3 \\
tempo_valid      & 5 &    1198704 & 0.93456 & 0.00388 & vgg_like_in=(40__256__1)_out=256_filters=16_pool_shape=(2__2)_max=True_filter_shapes=[(5__5)__(3__3)]_dropout=0.3 \\
tempo_valid      & 5 &    2668648 & 0.93250 & 0.00503 & vgg_like_in=(40__256__1)_out=256_filters=24_pool_shape=(2__2)_max=True_filter_shapes=[(5__5)__(3__3)]_dropout=0.5 \\
}\temposquarevgg


\pgfplotstableread[row sep=\\,col sep=&]{
Testset & Runs & Parameters & Acc & AccStd & Score & ScoreStd & Model \\
key_valid        & 5 &      12380 & 0.61442 & 0.01782 & 0.65709 & 0.01649 & shallow_key_in=(168__60__1)_out=24_filters=1_short_shape=(3__1)_long_shape=(168__1)_dropout=0.1 \\
key_valid        & 5 &      46240 & 0.62928 & 0.00979 & 0.67236 & 0.00994 & shallow_key_in=(168__60__1)_out=24_filters=2_short_shape=(3__1)_long_shape=(168__1)_dropout=0.3 \\
key_valid        & 5 &     178472 & 0.63724 & 0.00717 & 0.67955 & 0.00565 & shallow_key_in=(168__60__1)_out=24_filters=4_short_shape=(3__1)_long_shape=(168__1)_dropout=0.5 \\
key_valid        & 5 &     396720 & 0.63143 & 0.01112 & 0.67300 & 0.00965 & shallow_key_in=(168__60__1)_out=24_filters=6_short_shape=(3__1)_long_shape=(168__1)_dropout=0.3 \\
key_valid        & 5 &     700984 & 0.63208 & 0.00835 & 0.67576 & 0.00458 & shallow_key_in=(168__60__1)_out=24_filters=8_short_shape=(3__1)_long_shape=(168__1)_dropout=0.5 \\
key_valid        & 5 &    1567560 & 0.63488 & 0.00381 & 0.67572 & 0.00427 & shallow_key_in=(168__60__1)_out=24_filters=12_short_shape=(3__1)_long_shape=(168__1)_dropout=0.3 \\
}\keykeyshallow

\pgfplotstableread[row sep=\\,col sep=&]{
Testset & Runs & Parameters & Acc & AccStd & Score & ScoreStd & Model \\
key_valid        & 5 &       7004 & 0.56555 & 0.01153 & 0.61199 & 0.01204 & shallow_key_in=(168__60__1)_out=24_filters=1_short_shape=(3__1)_long_shape=(84__1)_dropout=0.1 \\
key_valid        & 5 &      24736 & 0.59354 & 0.00754 & 0.63869 & 0.00479 & shallow_key_in=(168__60__1)_out=24_filters=2_short_shape=(3__1)_long_shape=(84__1)_dropout=0.3 \\
key_valid        & 5 &      92456 & 0.60818 & 0.00326 & 0.65018 & 0.00511 & shallow_key_in=(168__60__1)_out=24_filters=4_short_shape=(3__1)_long_shape=(84__1)_dropout=0.3 \\
key_valid        & 5 &     203184 & 0.60603 & 0.00946 & 0.64964 & 0.00947 & shallow_key_in=(168__60__1)_out=24_filters=6_short_shape=(3__1)_long_shape=(84__1)_dropout=0.3 \\
key_valid        & 5 &     356920 & 0.61270 & 0.01693 & 0.65763 & 0.01618 & shallow_key_in=(168__60__1)_out=24_filters=8_short_shape=(3__1)_long_shape=(84__1)_dropout=0.1 \\
key_valid        & 5 &     793416 & 0.61033 & 0.01989 & 0.65356 & 0.01573 & shallow_key_in=(168__60__1)_out=24_filters=12_short_shape=(3__1)_long_shape=(84__1)_dropout=0.3 \\
}\keykeyshallowhalf

\pgfplotstableread[row sep=\\,col sep=&]{
Testset & Runs & Parameters & Acc & AccStd & Score & ScoreStd & Model \\
key_valid        & 5 &       5468 & 0.07298 & 0.01780 & 0.10868 & 0.02422 & shallow_tempo_in=(168__60__1)_out=24_filters=1_short_shape=(1__3)_long_shape=(1__60)_dropout=0.5 \\
key_valid        & 5 &      18592 & 0.06846 & 0.01636 & 0.10080 & 0.01913 & shallow_tempo_in=(168__60__1)_out=24_filters=2_short_shape=(1__3)_long_shape=(1__60)_dropout=0.5 \\
key_valid        & 5 &      67880 & 0.09107 & 0.03127 & 0.12446 & 0.02542 & shallow_tempo_in=(168__60__1)_out=24_filters=4_short_shape=(1__3)_long_shape=(1__60)_dropout=0.3 \\
key_valid        & 5 &     147888 & 0.06738 & 0.03714 & 0.09740 & 0.02937 & shallow_tempo_in=(168__60__1)_out=24_filters=6_short_shape=(1__3)_long_shape=(1__60)_dropout=0.3 \\
key_valid        & 5 &     258616 & 0.08245 & 0.01914 & 0.11326 & 0.01702 & shallow_tempo_in=(168__60__1)_out=24_filters=8_short_shape=(1__3)_long_shape=(1__60)_dropout=0.3 \\
key_valid        & 5 &     572232 & 0.07255 & 0.02487 & 0.10930 & 0.02897 & shallow_tempo_in=(168__60__1)_out=24_filters=12_short_shape=(1__3)_long_shape=(1__60)_dropout=0.1 \\
}\keytemposhallow

\pgfplotstableread[row sep=\\,col sep=&]{
Testset & Runs & Parameters & Acc & AccStd & Score & ScoreStd & Model \\
key_valid        & 5 &       3548 & 0.09494 & 0.01893 & 0.12762 & 0.01075 & shallow_tempo_in=(168__60__1)_out=24_filters=1_short_shape=(1__3)_long_shape=(1__30)_dropout=0.5 \\
key_valid        & 5 &      10912 & 0.07147 & 0.03948 & 0.11279 & 0.03643 & shallow_tempo_in=(168__60__1)_out=24_filters=2_short_shape=(1__3)_long_shape=(1__30)_dropout=0.5 \\
key_valid        & 5 &      37160 & 0.08267 & 0.02583 & 0.11324 & 0.02689 & shallow_tempo_in=(168__60__1)_out=24_filters=4_short_shape=(1__3)_long_shape=(1__30)_dropout=0.5 \\
key_valid        & 5 &      78768 & 0.08116 & 0.02539 & 0.12304 & 0.01877 & shallow_tempo_in=(168__60__1)_out=24_filters=6_short_shape=(1__3)_long_shape=(1__30)_dropout=0.5 \\
key_valid        & 5 &     135736 & 0.06545 & 0.04135 & 0.09572 & 0.02712 & shallow_tempo_in=(168__60__1)_out=24_filters=8_short_shape=(1__3)_long_shape=(1__30)_dropout=0.1 \\
key_valid        & 5 &     295752 & 0.09236 & 0.01846 & 0.12698 & 0.01602 & shallow_tempo_in=(168__60__1)_out=24_filters=12_short_shape=(1__3)_long_shape=(1__30)_dropout=0.3 \\
}\keytemposhallowhalf

\pgfplotstableread[row sep=\\,col sep=&]{
Testset & Runs & Parameters & Acc & AccStd & Score & ScoreStd & Model \\
key_valid        & 5 &       5378 & 0.64952 & 0.00530 & 0.68943 & 0.00333 & vgg_like_in=(168__60__1)_out=24_filters=2_pool_shape=(2__2)_max=True_filter_shapes=[(5__1)__(3__1)]_dropout=0.1 \\
key_valid        & 5 &      19572 & 0.66243 & 0.00827 & 0.70243 & 0.00734 & vgg_like_in=(168__60__1)_out=24_filters=4_pool_shape=(2__2)_max=True_filter_shapes=[(5__1)__(3__1)]_dropout=0.1 \\
key_valid        & 5 &      74480 & 0.67449 & 0.00907 & 0.71376 & 0.00841 & vgg_like_in=(168__60__1)_out=24_filters=8_pool_shape=(2__2)_max=True_filter_shapes=[(5__1)__(3__1)]_dropout=0.1 \\
key_valid        & 5 &     290376 & 0.68332 & 0.00651 & 0.72084 & 0.00583 & vgg_like_in=(168__60__1)_out=24_filters=16_pool_shape=(2__2)_max=True_filter_shapes=[(5__1)__(3__1)]_dropout=0.3 \\
key_valid        & 5 &     647712 & 0.68913 & 0.00433 & 0.72543 & 0.00355 & vgg_like_in=(168__60__1)_out=24_filters=24_pool_shape=(2__2)_max=True_filter_shapes=[(5__1)__(3__1)]_dropout=0.3 \\
}\keykeyvgg

\pgfplotstableread[row sep=\\,col sep=&]{
Testset & Runs & Parameters & Acc & AccStd & Score & ScoreStd & Model \\
key_valid        & 5 &       5378 & 0.14058 & 0.02537 & 0.17765 & 0.02094 & vgg_like_in=(168__60__1)_out=24_filters=2_pool_shape=(2__2)_max=True_filter_shapes=[(1__5)__(1__3)]_dropout=0.1 \\
key_valid        & 5 &      19572 & 0.33864 & 0.06213 & 0.39122 & 0.06126 & vgg_like_in=(168__60__1)_out=24_filters=4_pool_shape=(2__2)_max=True_filter_shapes=[(1__5)__(1__3)]_dropout=0.1 \\
key_valid        & 5 &      74480 & 0.49795 & 0.01165 & 0.54235 & 0.00996 & vgg_like_in=(168__60__1)_out=24_filters=8_pool_shape=(2__2)_max=True_filter_shapes=[(1__5)__(1__3)]_dropout=0.1 \\
key_valid        & 5 &     290376 & 0.54661 & 0.01902 & 0.59191 & 0.01715 & vgg_like_in=(168__60__1)_out=24_filters=16_pool_shape=(2__2)_max=True_filter_shapes=[(1__5)__(1__3)]_dropout=0.1 \\
key_valid        & 5 &     647712 & 0.57567 & 0.00883 & 0.61929 & 0.00720 & vgg_like_in=(168__60__1)_out=24_filters=24_pool_shape=(2__2)_max=True_filter_shapes=[(1__5)__(1__3)]_dropout=0.3 \\
}\keytempovgg

\pgfplotstableread[row sep=\\,col sep=&]{
Testset & Runs & Parameters & Acc & AccStd & Score & ScoreStd & Model \\
key_valid        & 5 &       5046 & 0.42110 & 0.06938 & 0.47686 & 0.05821 & vgg_like_in=(168__60__1)_out=24_filters=1_pool_shape=(2__2)_max=True_filter_shapes=[(5__5)__(3__3)]_dropout=0.1 \\
key_valid        & 5 &      19138 & 0.61615 & 0.01376 & 0.66263 & 0.01511 & vgg_like_in=(168__60__1)_out=24_filters=2_pool_shape=(2__2)_max=True_filter_shapes=[(5__5)__(3__3)]_dropout=0.3 \\
key_valid        & 5 &      74532 & 0.65339 & 0.00960 & 0.69259 & 0.00992 & vgg_like_in=(168__60__1)_out=24_filters=4_pool_shape=(2__2)_max=True_filter_shapes=[(5__5)__(3__3)]_dropout=0.3 \\
key_valid        & 5 &     294160 & 0.67040 & 0.00591 & 0.70827 & 0.00457 & vgg_like_in=(168__60__1)_out=24_filters=8_pool_shape=(2__2)_max=True_filter_shapes=[(5__5)__(3__3)]_dropout=0.1 \\
key_valid        & 5 &    1168776 & 0.67018 & 0.01124 & 0.71091 & 0.00780 & vgg_like_in=(168__60__1)_out=24_filters=16_pool_shape=(2__2)_max=True_filter_shapes=[(5__5)__(3__3)]_dropout=0.5 \\
key_valid        & 5 &    2623872 & 0.67384 & 0.01627 & 0.71076 & 0.01226 & vgg_like_in=(168__60__1)_out=24_filters=24_pool_shape=(2__2)_max=True_filter_shapes=[(5__5)__(3__3)]_dropout=0.5 \\
}\keysquarevgg

\begin{figure}[t]
\centering
\begin{tikzpicture} 
\begin{axis}[
            name=ax1,
            title=Tempo Validation,
            width=4.5cm,
            height=4.5cm,
			grid=both,
			ymin=0,
			ymax=1,
			xmin=4000,
			xmax=3500000,
			xmode=log,
			tick align=outside,
            ylabel={Accuracy in \%},
            xlabel={Parameters},
            yticklabel=\pgfmathparse{100*\tick}\pgfmathprintnumber{\pgfmathresult},
]
    \addplot[red, mark=x, error bars/.cd, y dir = both, y explicit] table[x=Parameters,y=Acc1,y error=Std1]{\tempokeyshallow};
    \addplot[blue, mark=x, error bars/.cd, y dir = both, y explicit] table[x=Parameters,y=Acc1,y error=Std1]{\tempotemposhallow};
    \addplot[red, mark=none, error bars/.cd, y dir = both, y explicit] table[x=Parameters,y=Acc1,y error=Std1]{\tempokeyvgg};
    \addplot[blue, mark=none, error bars/.cd, y dir = both, y explicit] table[x=Parameters,y=Acc1,y error=Std1]{\tempotempovgg};
    \addplot[black!90!white, mark=none, error bars/.cd, y dir = both, y explicit] table[x=Parameters,y=Acc1,y error=Std1]{\temposquarevgg};
	\addplot[thick, dashed, samples=50, smooth, black!90!white] coordinates{(1000,0.038)(10000000,0.038)};
	\addplot[thick, dotted, samples=50, smooth, black!90!white] coordinates{(1000,0.287)(10000000,0.287)};
\end{axis} 
\begin{axis}[
            title=Key Validation,
            at={(ax1.south east)},
	        xshift=0.25cm,
            width=4.5cm,
            height=4.5cm,
			grid=both,
			xmode=log,
			ymin=0,
			ymax=1,
			xmin=4000,
			xmax=3500000,
			tick align=outside,
            legend style={at={(-0.1,-0.38)},anchor=north},
            legend columns=2,
            legend style={draw=none,font=\small},
            legend cell align={left},
            xlabel={Parameters},
            yticklabels={,,}
]
    \addplot[red, mark=x, error bars/.cd, y dir = both, y explicit] table[x=Parameters,y=Acc,y error=AccStd]{\keykeyshallow};
    \addplot[blue, mark=x, error bars/.cd, y dir = both, y explicit] table[x=Parameters,y=Acc,y error=AccStd]{\keytemposhallow};
    \addplot[red, mark=none, error bars/.cd, y dir = both, y explicit] table[x=Parameters,y=Acc,y error=AccStd]{\keykeyvgg};
    \addplot[blue, mark=none, error bars/.cd, y dir = both, y explicit] table[x=Parameters,y=Acc,y error=AccStd]{\keytempovgg};
    \addplot[black!90!white, mark=none, error bars/.cd, y dir = both, y explicit] table[x=Parameters,y=Acc,y error=AccStd]{\keysquarevgg};
    \addlegendentry{\shallowkey}
    \addlegendentry{\shallowtempo}
    \addlegendentry{\deepkey}
    \addlegendentry{\deeptempo}
    \addlegendentry{\deepsquare}
	\addplot[thick, dashed, samples=50, smooth, black!90!white] coordinates{(1000,0.0417)(10000000,0.0417)};
	\addplot[thick, dotted, samples=50, smooth, black!90!white] coordinates{(1000,0.141)(10000000,0.141)};
\end{axis} 

\end{tikzpicture}

\caption{
Mean validation accuracies for the various network configurations depending on
their number of network parameters.
Only the best dropout configuration is shown.
Whiskers represent the standard deviation based on 5 runs.
}
\label{fig:validation}
\end{figure}
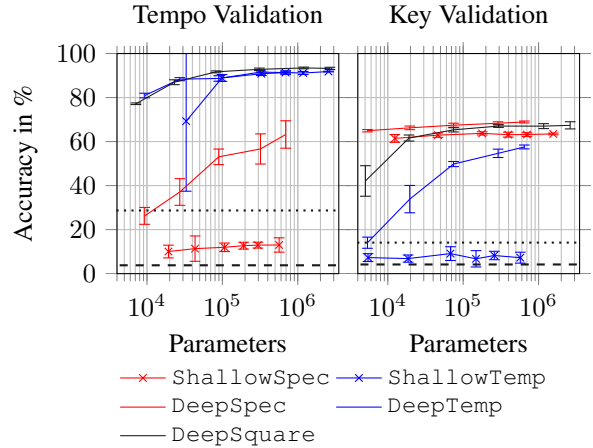

\figref{fig:validation} shows mean validation
accuracies of 5 runs for each configuration, using their best performing dropout
probability $p_\mathrm{D}$. The dashed black line is the accuracy a
random classifier achieves, and the dotted black line shows the accuracy
of the algorithm that always outputs the class that most often occurs
in the validation set.
With accuracy values slightly above random, \shallowkey\ and \shallowtempo\ perform
worst of all architectures, when used for the task they were \emph{not} meant for.
But when used for the task they were designed for, both perform well.
A higher number of parameters leads to slightly better results.
When training \shallowtempo\ with $k=1$ for the tempo task, the network performed
very poorly in one of the five runs, which is the cause for the very large standard deviation
of $32.2$ shown in \figref{fig:validation}. The mean accuracy for the 4
successful runs was $85.2\%$.
When comparing with the \deep\ architectures, we see that \deeptempo\ performs just
as well as \shallowtempo\ with $k>1$ on the tempo task, and \deepkey\ clearly outperforms
\shallowkey\ on the key task.
Surprisingly, the \deepkey\ architecture reaches fairly high accuracy
values (up~to~63\%) on the tempo task when increasing the model capacity via $k$,
even though it only has convolutional filters aligned with the frequency axis.
We can make a similar observation for the \deeptempo\ architecture. It too reaches
relatively high accuracy values on the key task (up~to~57\%) when increasing $k$.
The unspecialized \deepsquare\ is by a small margin the best
performing architecture for the tempo task,
and comes in as a close second for key detection with $k>1$.
But for $k=1$, \deepsquare\ performs considerably worse than
\deepkey\ with $k=2$~($42\%$ compared to $64\%$), even though both have similar
parameter counts of ca.\ $5\,000$.

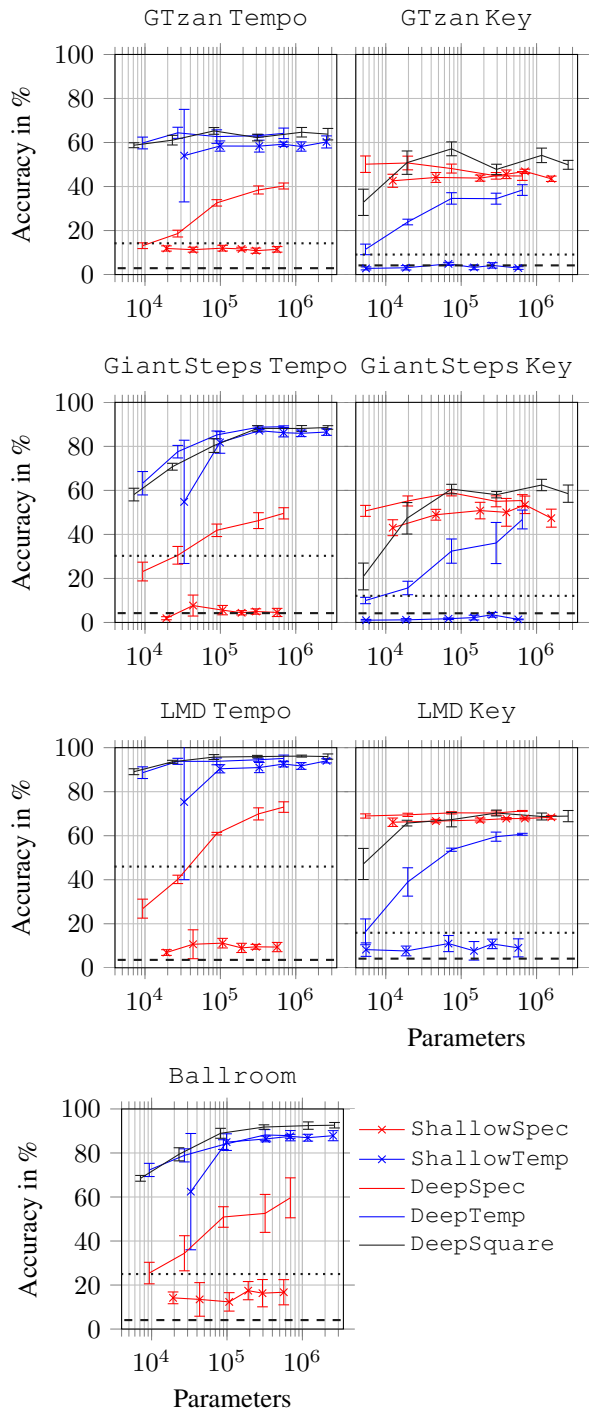
\begin{figure}[!ht]
\centering
\begin{tikzpicture} 
\begin{axis}[
            name=ax1,
            title=\gtzantempo,
            width=4.5cm,
            height=4.5cm,
			grid=both,
			ymin=0,
			ymax=1,
			xmin=4000,
			xmax=3500000,
			xmode=log,
			tick align=outside,
            ylabel={Accuracy in \%},
            yticklabel=\pgfmathparse{100*\tick}\pgfmathprintnumber{\pgfmathresult},
]
    \addplot[red, mark=x, error bars/.cd, y dir = both, y explicit] table[x=Parameters,y=Acc1,y error=Std1]{\gtzantempokeyshallow};
    \addplot[blue, mark=x, error bars/.cd, y dir = both, y explicit] table[x=Parameters,y=Acc1,y error=Std1]{\gtzantempotemposhallow};
    \addplot[red, mark=none, error bars/.cd, y dir = both, y explicit] table[x=Parameters,y=Acc1,y error=Std1]{\gtzantempokeyvgg};
    \addplot[blue, mark=none, error bars/.cd, y dir = both, y explicit] table[x=Parameters,y=Acc1,y error=Std1]{\gtzantempotempovgg};
    \addplot[black!90!white, mark=none, error bars/.cd, y dir = both, y explicit] table[x=Parameters,y=Acc1,y error=Std1]{\gtzantemposquarevgg};
	\addplot[thick, dashed, samples=50, smooth, black!90!white] coordinates{(1000,0.029)(10000000,0.029)};
	\addplot[thick, dotted, samples=50, smooth, black!90!white] coordinates{(1000,0.142)(10000000,0.142)};
\end{axis} 
\begin{axis}[
            title=\gtzankey,
            at={(ax1.south east)},
	        xshift=0.25cm,
            width=4.5cm,
            height=4.5cm,
			grid=both,
			xmode=log,
			ymin=0,
			ymax=1,
			xmin=4000,
			xmax=3500000,
			tick align=outside,
            legend style={at={(-0.1,-0.38)},anchor=north},
            legend columns=2,
            legend style={draw=none},
            legend cell align={left},
            yticklabels={,,}
]
    \addplot[red, mark=x, error bars/.cd, y dir = both, y explicit] table[x=Parameters,y=Acc,y error=AccStd]{\gtzankeykeyshallow};
    \addplot[blue, mark=x, error bars/.cd, y dir = both, y explicit] table[x=Parameters,y=Acc,y error=AccStd]{\gtzankeytemposhallow};
    \addplot[red, mark=none, error bars/.cd, y dir = both, y explicit] table[x=Parameters,y=Acc,y error=AccStd]{\gtzankeykeyvgg};
    \addplot[blue, mark=none, error bars/.cd, y dir = both, y explicit] table[x=Parameters,y=Acc,y error=AccStd]{\gtzankeytempovgg};
    \addplot[black!90!white, mark=none, error bars/.cd, y dir = both, y explicit] table[x=Parameters,y=Acc,y error=AccStd]{\gtzankeysquarevgg};
	\addplot[thick, dashed, samples=50, smooth, black!90!white] coordinates{(1000,0.0417)(10000000,0.0417)};
	\addplot[thick, dotted, samples=50, smooth, black!90!white] coordinates{(1000,0.0909)(10000000,0.0909)};
\end{axis} 

\end{tikzpicture}
%
%
\begin{tikzpicture} 
\begin{axis}[
            name=ax1,
            title=\giantstepstempo,
            width=4.5cm,
            height=4.5cm,
			grid=both,
			ymin=0,
			ymax=1,
			xmin=4000,
			xmax=3500000,
			xmode=log,
			tick align=outside,
            ylabel={Accuracy in \%},
            yticklabel=\pgfmathparse{100*\tick}\pgfmathprintnumber{\pgfmathresult},
]
    \addplot[red, mark=x, error bars/.cd, y dir = both, y explicit] table[x=Parameters,y=Acc1,y error=Std1]{\giantstepstempokeyshallow};
    \addplot[blue, mark=x, error bars/.cd, y dir = both, y explicit] table[x=Parameters,y=Acc1,y error=Std1]{\giantstepstempotemposhallow};
    \addplot[red, mark=none, error bars/.cd, y dir = both, y explicit] table[x=Parameters,y=Acc1,y error=Std1]{\giantstepstempokeyvgg};
    \addplot[blue, mark=none, error bars/.cd, y dir = both, y explicit] table[x=Parameters,y=Acc1,y error=Std1]{\giantstepstempotempovgg};
    \addplot[black!90!white, mark=none, error bars/.cd, y dir = both, y explicit] table[x=Parameters,y=Acc1,y error=Std1]{\giantstepstemposquarevgg};
	\addplot[thick, dashed, samples=50, smooth, black!90!white] coordinates{(1000,0.0424)(10000000,0.0424)};
	\addplot[thick, dotted, samples=50, smooth, black!90!white] coordinates{(1000,0.3027)(10000000,0.3027)};
\end{axis} 
\begin{axis}[
            title=\giantstepskey,
            at={(ax1.south east)},
	        xshift=0.25cm,
            width=4.5cm,
            height=4.5cm,
			grid=both,
			xmode=log,
			ymin=0,
			ymax=1,
			xmin=4000,
			xmax=3500000,
			tick align=outside,
            legend style={at={(-0.1,-0.38)},anchor=north},
            legend columns=2,
            legend style={draw=none},
            legend cell align={left},
            yticklabels={,,}
]
    \addplot[red, mark=x, error bars/.cd, y dir = both, y explicit] table[x=Parameters,y=Acc,y error=AccStd]{\giantstepskeykeyshallow};
    \addplot[blue, mark=x, error bars/.cd, y dir = both, y explicit] table[x=Parameters,y=Acc,y error=AccStd]{\giantstepskeytemposhallow};
    \addplot[red, mark=none, error bars/.cd, y dir = both, y explicit] table[x=Parameters,y=Acc,y error=AccStd]{\giantstepskeykeyvgg};
    \addplot[blue, mark=none, error bars/.cd, y dir = both, y explicit] table[x=Parameters,y=Acc,y error=AccStd]{\giantstepskeytempovgg};
    \addplot[black!90!white, mark=none, error bars/.cd, y dir = both, y explicit] table[x=Parameters,y=Acc,y error=AccStd]{\giantstepskeysquarevgg};
	\addplot[thick, dashed, samples=50, smooth, black!90!white] coordinates{(1000,0.04169)(10000000,0.04169)};
	\addplot[thick, dotted, samples=50, smooth, black!90!white] coordinates{(1000,0.12086)(10000000,0.12086)};
\end{axis} 

\end{tikzpicture}
%
%
%
\begin{tikzpicture} 
\begin{axis}[
            name=ax1,
            title=\lmdtempo,
            width=4.5cm,
            height=4.5cm,
			grid=both,
			ymin=0,
			ymax=1,
			xmin=4000,
			xmax=3500000,
			xmode=log,
			tick align=outside,
            ylabel={Accuracy in \%},
            yticklabel=\pgfmathparse{100*\tick}\pgfmathprintnumber{\pgfmathresult},
]
    \addplot[red, mark=x, error bars/.cd, y dir = both, y explicit] table[x=Parameters,y=Acc1,y error=Std1]{\lmdtempokeyshallow};
    \addplot[blue, mark=x, error bars/.cd, y dir = both, y explicit] table[x=Parameters,y=Acc1,y error=Std1]{\lmdtempotemposhallow};
    \addplot[red, mark=none, error bars/.cd, y dir = both, y explicit] table[x=Parameters,y=Acc1,y error=Std1]{\lmdtempokeyvgg};
    \addplot[blue, mark=none, error bars/.cd, y dir = both, y explicit] table[x=Parameters,y=Acc1,y error=Std1]{\lmdtempotempovgg};
    \addplot[black!90!white, mark=none, error bars/.cd, y dir = both, y explicit] table[x=Parameters,y=Acc1,y error=Std1]{\lmdtemposquarevgg};
	\addplot[thick, dashed, samples=50, smooth, black!90!white] coordinates{(1000,0.0359)(10000000,0.0359)};
	\addplot[thick, dotted, samples=50, smooth, black!90!white] coordinates{(1000,0.4598)(10000000,0.4598)};
\end{axis} 
\begin{axis}[
            title=\lmdkey,
            at={(ax1.south east)},
	        xshift=0.25cm,
            width=4.5cm,
            height=4.5cm,
			grid=both,
			xmode=log,
			ymin=0,
			ymax=1,
			xmin=4000,
			xmax=3500000,
			tick align=outside,
            legend style={at={(-0.1,-0.38)},anchor=north},
            legend columns=2,
            legend style={draw=none},
            legend cell align={left},
            xlabel={Parameters},
            yticklabels={,,}
]
    \addplot[red, mark=x, error bars/.cd, y dir = both, y explicit] table[x=Parameters,y=Acc,y error=AccStd]{\lmdkeykeyshallow};
    \addplot[blue, mark=x, error bars/.cd, y dir = both, y explicit] table[x=Parameters,y=Acc,y error=AccStd]{\lmdkeytemposhallow};
    \addplot[red, mark=none, error bars/.cd, y dir = both, y explicit] table[x=Parameters,y=Acc,y error=AccStd]{\lmdkeykeyvgg};
    \addplot[blue, mark=none, error bars/.cd, y dir = both, y explicit] table[x=Parameters,y=Acc,y error=AccStd]{\lmdkeytempovgg};
    \addplot[black!90!white, mark=none, error bars/.cd, y dir = both, y explicit] table[x=Parameters,y=Acc,y error=AccStd]{\lmdkeysquarevgg};
	\addplot[thick, dashed, samples=50, smooth, black!90!white] coordinates{(1000,0.04152)(10000000,0.04152)};
	\addplot[thick, dotted, samples=50, smooth, black!90!white] coordinates{(1000,0.1590)(10000000,0.1590)};
\end{axis} 

\end{tikzpicture}
%
\begin{tikzpicture} 
\begin{axis}[
            name=ax1,
            title=\ballroom,
            yshift=-3cm,
            width=4.5cm,
            height=4.5cm,
			grid=both,
			ymin=0,
			ymax=1,
			xmin=4000,
			xmax=3500000,
			xmode=log,
			tick align=outside,
            legend style={draw=none,font=\small},
            legend pos=outer north east,
            legend cell align={left},
            ylabel={Accuracy in \%},
            xlabel={Parameters},
            yticklabel=\pgfmathparse{100*\tick}\pgfmathprintnumber{\pgfmathresult},
]
    \addplot[red, mark=x, error bars/.cd, y dir = both, y explicit] table[x=Parameters,y=Acc1,y error=Std1]{\ballroomtempokeyshallow};
    \addplot[blue, mark=x, error bars/.cd, y dir = both, y explicit] table[x=Parameters,y=Acc1,y error=Std1]{\ballroomtempotemposhallow};
    \addplot[red, mark=none, error bars/.cd, y dir = both, y explicit] table[x=Parameters,y=Acc1,y error=Std1]{\ballroomtempokeyvgg};
    \addplot[blue, mark=none, error bars/.cd, y dir = both, y explicit] table[x=Parameters,y=Acc1,y error=Std1]{\ballroomtempotempovgg};
    \addplot[black!90!white, mark=none, error bars/.cd, y dir = both, y explicit] table[x=Parameters,y=Acc1,y error=Std1]{\ballroomtemposquarevgg};
    \addlegendentry{\shallowkey}
    \addlegendentry{\shallowtempo}
    \addlegendentry{\deepkey}
    \addlegendentry{\deeptempo}
    \addlegendentry{\deepsquare}
	\addplot[thick, dashed, samples=50, smooth, black!90!white] coordinates{(1000,0.0408)(10000000,0.0408)};
	\addplot[thick, dotted, samples=50, smooth, black!90!white] coordinates{(1000,0.25)(10000000,0.25)};
\end{axis} 
\end{tikzpicture}
\caption{Mean test accuracies for various network configurations and datasets depending on their
number of network parameters. Whiskers represent one standard deviation based on 5 runs.
Dropout was chosen based on performance during validation.
}
\label{fig:test}
\end{figure}

We selected the dropout variant for each architecture and parameter setting with
the best validation accuracy and ran predictions on the test sets.
Detailed results are shown in \figref{fig:test}.
The general picture is very similar to validation:
\deep\ architectures tend to perform slightly better than \shallow\ architectures
on the tasks they are meant for and \shallow\ architectures perform poorly on
the task they were not meant for. In fact, \shallowtempo\ performs no better on
\gtzankey\ and \giantstepskey\ than the random baseline. For both key and tempo
\deepsquare\ performs as well or better than any other architecture, except when
drastically reducing the model capacity for the key task~($k=1$).
Then accuracy decreases well below \deepkey's performance with similar parameter
count: $33\%$ compared to $50\%$ for \gtzankey, and $21\%$ compared to $51\%$ for
\giantstepskey.

To provide an absolute comparison, we chose the best performing representative from
each architecture (based on validation accuracy, regardless of dropout configuration
or capacity), and calculated accuracies for each test set (\tabref{tab:test}, incl.\
reference values from the literature).
In 5 out of 7 test cases \deepsquare\ reaches the highest accuracy score among our
architectures. The other two are reached by \deeptempo\ for \giantstepstempo\
and by \deepkey\ for \lmdkey.
For both tasks we observe that the margin by which the best performing network
is better than the second best for a given dataset differs considerably.
\deepsquare\ reaches an accuracy of $92.4\%$ for the \ballroom\ tempo dataset, which
is $4.2\,\mathrm{pp}$ (percentage points) better than the second best network (\deeptempo, $88.2\%$).
The differences between best and second best
accuracy are considerably lower for the other datasets:
$1.7\,\mathrm{pp}$ (\lmdtempo),
$1.6\,\mathrm{pp}$ (\gtzantempo), and
$0.6\,\mathrm{pp}$ (\giantstepstempo).
For the key task, \deepsquare\ reaches an accuracy of $49.9\%$ on \gtzankey, which
is $5.1\,\mathrm{pp}$ better than the second best network (\deepkey, $44.8\%$),
while the differences between best and second best for the other datasets are 
$3.1\,\mathrm{pp}$ (\giantstepskey), and
$2.4\,\mathrm{pp}$ (\lmdkey).


\pgfplotstableread[row sep=\\,col sep=&]{
Testset & Runs & Parameters & Acc1 & Std1 & Model \\
gtzan            & 5 &      19268 & 0.11852 & 0.01189 & shallow_key_in=(40__256__1)_out=256_filters=1_short_shape=(3__1)_long_shape=(40__1)_dropout=0.1 \\
gtzan            & 5 &      43400 & 0.11311 & 0.01129 & shallow_key_in=(40__256__1)_out=256_filters=2_short_shape=(3__1)_long_shape=(40__1)_dropout=0.3 \\
gtzan            & 5 &     107024 & 0.11992 & 0.01298 & shallow_key_in=(40__256__1)_out=256_filters=4_short_shape=(3__1)_long_shape=(40__1)_dropout=0.3 \\
gtzan            & 5 &     191128 & 0.11632 & 0.00797 & shallow_key_in=(40__256__1)_out=256_filters=6_short_shape=(3__1)_long_shape=(40__1)_dropout=0.3 \\
gtzan            & 5 &     295712 & 0.10851 & 0.01189 & shallow_key_in=(40__256__1)_out=256_filters=8_short_shape=(3__1)_long_shape=(40__1)_dropout=0.1 \\
gtzan            & 5 &     566320 & 0.11451 & 0.01258 & shallow_key_in=(40__256__1)_out=256_filters=12_short_shape=(3__1)_long_shape=(40__1)_dropout=0.5 \\
}\gtzantempokeyshallow

\pgfplotstableread[row sep=\\,col sep=&]{
Testset & Runs & Parameters & Acc1 & Std1 & Model \\
gtzan            & 5 &      17988 & 0.11371 & 0.00524 & shallow_key_in=(40__256__1)_out=256_filters=1_short_shape=(3__1)_long_shape=(20__1)_dropout=0.3 \\
gtzan            & 5 &      38280 & 0.10951 & 0.00959 & shallow_key_in=(40__256__1)_out=256_filters=2_short_shape=(3__1)_long_shape=(20__1)_dropout=0.3 \\
gtzan            & 5 &      86544 & 0.11471 & 0.00764 & shallow_key_in=(40__256__1)_out=256_filters=4_short_shape=(3__1)_long_shape=(20__1)_dropout=0.3 \\
gtzan            & 5 &     145048 & 0.11251 & 0.01730 & shallow_key_in=(40__256__1)_out=256_filters=6_short_shape=(3__1)_long_shape=(20__1)_dropout=0.1 \\
gtzan            & 5 &     213792 & 0.10651 & 0.00698 & shallow_key_in=(40__256__1)_out=256_filters=8_short_shape=(3__1)_long_shape=(20__1)_dropout=0.5 \\
gtzan            & 5 &     382000 & 0.10290 & 0.01128 & shallow_key_in=(40__256__1)_out=256_filters=12_short_shape=(3__1)_long_shape=(20__1)_dropout=0.5 \\
}\gtzantempokeyshallowhalf

\pgfplotstableread[row sep=\\,col sep=&]{
Testset & Runs & Parameters & Acc1 & Std1 & Model \\
gtzan            & 5 &      33092 & 0.54034 & 0.21020 & shallow_tempo_in=(40__256__1)_out=256_filters=1_short_shape=(1__3)_long_shape=(1__256)_dropout=0.1 \\
gtzan            & 5 &      98696 & 0.58398 & 0.02321 & shallow_tempo_in=(40__256__1)_out=256_filters=2_short_shape=(1__3)_long_shape=(1__256)_dropout=0.3 \\
gtzan            & 5 &     328208 & 0.58358 & 0.02701 & shallow_tempo_in=(40__256__1)_out=256_filters=4_short_shape=(1__3)_long_shape=(1__256)_dropout=0.5 \\
gtzan            & 5 &     688792 & 0.59219 & 0.00980 & shallow_tempo_in=(40__256__1)_out=256_filters=6_short_shape=(1__3)_long_shape=(1__256)_dropout=0.5 \\
gtzan            & 5 &    1180448 & 0.58178 & 0.02113 & shallow_tempo_in=(40__256__1)_out=256_filters=8_short_shape=(1__3)_long_shape=(1__256)_dropout=0.3 \\
gtzan            & 5 &    2556976 & 0.60260 & 0.02747 & shallow_tempo_in=(40__256__1)_out=256_filters=12_short_shape=(1__3)_long_shape=(1__256)_dropout=0.5 \\
}\gtzantempotemposhallow

\pgfplotstableread[row sep=\\,col sep=&]{
Testset & Runs & Parameters & Acc1 & Std1 & Model \\
gtzan            & 5 &      24900 & 0.54535 & 0.21447 & shallow_tempo_in=(40__256__1)_out=256_filters=1_short_shape=(1__3)_long_shape=(1__128)_dropout=0.3 \\
gtzan            & 5 &      65928 & 0.65085 & 0.02276 & shallow_tempo_in=(40__256__1)_out=256_filters=2_short_shape=(1__3)_long_shape=(1__128)_dropout=0.5 \\
gtzan            & 5 &     197136 & 0.58478 & 0.02451 & shallow_tempo_in=(40__256__1)_out=256_filters=4_short_shape=(1__3)_long_shape=(1__128)_dropout=0.3 \\
gtzan            & 5 &     393880 & 0.58999 & 0.02535 & shallow_tempo_in=(40__256__1)_out=256_filters=6_short_shape=(1__3)_long_shape=(1__128)_dropout=0.5 \\
gtzan            & 5 &     656160 & 0.57017 & 0.01212 & shallow_tempo_in=(40__256__1)_out=256_filters=8_short_shape=(1__3)_long_shape=(1__128)_dropout=0.3 \\
gtzan            & 5 &    1377328 & 0.59259 & 0.01229 & shallow_tempo_in=(40__256__1)_out=256_filters=12_short_shape=(1__3)_long_shape=(1__128)_dropout=0.5 \\
}\gtzantempotemposhallowhalf

\pgfplotstableread[row sep=\\,col sep=&]{
Testset & Runs & Parameters & Acc1 & Std1 & Model \\
gtzan            & 5 &       9322 & 0.59780 & 0.02674 & vgg_like_in=(40__256__1)_out=256_filters=2_pool_shape=(2__2)_max=True_filter_shapes=[(1__5)__(1__3)]_dropout=0.1 \\
gtzan            & 5 &      27228 & 0.64404 & 0.02524 & vgg_like_in=(40__256__1)_out=256_filters=4_pool_shape=(2__2)_max=True_filter_shapes=[(1__5)__(1__3)]_dropout=0.1 \\
gtzan            & 5 &      89560 & 0.62683 & 0.03067 & vgg_like_in=(40__256__1)_out=256_filters=8_pool_shape=(2__2)_max=True_filter_shapes=[(1__5)__(1__3)]_dropout=0.1 \\
gtzan            & 5 &     320304 & 0.63103 & 0.00554 & vgg_like_in=(40__256__1)_out=256_filters=16_pool_shape=(2__2)_max=True_filter_shapes=[(1__5)__(1__3)]_dropout=0.1 \\
gtzan            & 5 &     692488 & 0.64164 & 0.02390 & vgg_like_in=(40__256__1)_out=256_filters=24_pool_shape=(2__2)_max=True_filter_shapes=[(1__5)__(1__3)]_dropout=0.3 \\
}\gtzantempotempovgg

\pgfplotstableread[row sep=\\,col sep=&]{
Testset & Runs & Parameters & Acc1 & Std1 & Model \\
gtzan            & 5 &       9322 & 0.13113 & 0.01310 & vgg_like_in=(40__256__1)_out=256_filters=2_pool_shape=(2__2)_max=True_filter_shapes=[(5__1)__(3__1)]_dropout=0.1 \\
gtzan            & 5 &      27228 & 0.18619 & 0.01540 & vgg_like_in=(40__256__1)_out=256_filters=4_pool_shape=(2__2)_max=True_filter_shapes=[(5__1)__(3__1)]_dropout=0.1 \\
gtzan            & 5 &      89560 & 0.32573 & 0.01451 & vgg_like_in=(40__256__1)_out=256_filters=8_pool_shape=(2__2)_max=True_filter_shapes=[(5__1)__(3__1)]_dropout=0.1 \\
gtzan            & 5 &     320304 & 0.38438 & 0.01830 & vgg_like_in=(40__256__1)_out=256_filters=16_pool_shape=(2__2)_max=True_filter_shapes=[(5__1)__(3__1)]_dropout=0.1 \\
gtzan            & 5 &     692488 & 0.40240 & 0.01365 & vgg_like_in=(40__256__1)_out=256_filters=24_pool_shape=(2__2)_max=True_filter_shapes=[(5__1)__(3__1)]_dropout=0.1 \\
}\gtzantempokeyvgg

\pgfplotstableread[row sep=\\,col sep=&]{
Testset & Runs & Parameters & Acc1 & Std1 & Model \\
gtzan            & 5 &       7134 & 0.58759 & 0.01120 & vgg_like_in=(40__256__1)_out=256_filters=1_pool_shape=(2__2)_max=True_filter_shapes=[(5__5)__(3__3)]_dropout=0.1 \\
gtzan            & 5 &      23082 & 0.61061 & 0.02194 & vgg_like_in=(40__256__1)_out=256_filters=2_pool_shape=(2__2)_max=True_filter_shapes=[(5__5)__(3__3)]_dropout=0.1 \\
gtzan            & 5 &      82188 & 0.65265 & 0.01529 & vgg_like_in=(40__256__1)_out=256_filters=4_pool_shape=(2__2)_max=True_filter_shapes=[(5__5)__(3__3)]_dropout=0.1 \\
gtzan            & 5 &     309240 & 0.62242 & 0.01508 & vgg_like_in=(40__256__1)_out=256_filters=8_pool_shape=(2__2)_max=True_filter_shapes=[(5__5)__(3__3)]_dropout=0.3 \\
gtzan            & 5 &    1198704 & 0.64665 & 0.02161 & vgg_like_in=(40__256__1)_out=256_filters=16_pool_shape=(2__2)_max=True_filter_shapes=[(5__5)__(3__3)]_dropout=0.3 \\
gtzan            & 5 &    2668648 & 0.63784 & 0.02604 & vgg_like_in=(40__256__1)_out=256_filters=24_pool_shape=(2__2)_max=True_filter_shapes=[(5__5)__(3__3)]_dropout=0.5 \\
}\gtzantemposquarevgg


\pgfplotstableread[row sep=\\,col sep=&]{
Testset & Runs & Parameters & Acc & AccStd & Score & ScoreStd & Model \\
gtzan_key        & 5 &      12380 & 0.42608 & 0.02969 & 0.49818 & 0.02598 & shallow_key_in=(168__60__1)_out=24_filters=1_short_shape=(3__1)_long_shape=(168__1)_dropout=0.1 \\
gtzan_key        & 5 &      46240 & 0.44115 & 0.02240 & 0.51349 & 0.01987 & shallow_key_in=(168__60__1)_out=24_filters=2_short_shape=(3__1)_long_shape=(168__1)_dropout=0.3 \\
gtzan_key        & 5 &     178472 & 0.43828 & 0.01392 & 0.51333 & 0.01018 & shallow_key_in=(168__60__1)_out=24_filters=4_short_shape=(3__1)_long_shape=(168__1)_dropout=0.5 \\
gtzan_key        & 5 &     396720 & 0.45550 & 0.01957 & 0.52699 & 0.01535 & shallow_key_in=(168__60__1)_out=24_filters=6_short_shape=(3__1)_long_shape=(168__1)_dropout=0.3 \\
gtzan_key        & 5 &     700984 & 0.46890 & 0.00972 & 0.53856 & 0.00858 & shallow_key_in=(168__60__1)_out=24_filters=8_short_shape=(3__1)_long_shape=(168__1)_dropout=0.5 \\
gtzan_key        & 5 &    1567560 & 0.43517 & 0.01218 & 0.50974 & 0.01226 & shallow_key_in=(168__60__1)_out=24_filters=12_short_shape=(3__1)_long_shape=(168__1)_dropout=0.3 \\
}\gtzankeykeyshallow

\pgfplotstableread[row sep=\\,col sep=&]{
Testset & Runs & Parameters & Acc & AccStd & Score & ScoreStd & Model \\
gtzan_key        & 5 &       7004 & 0.37679 & 0.01825 & 0.45533 & 0.01515 & shallow_key_in=(168__60__1)_out=24_filters=1_short_shape=(3__1)_long_shape=(84__1)_dropout=0.1 \\
gtzan_key        & 5 &      24736 & 0.40670 & 0.00655 & 0.47876 & 0.00692 & shallow_key_in=(168__60__1)_out=24_filters=2_short_shape=(3__1)_long_shape=(84__1)_dropout=0.3 \\
gtzan_key        & 5 &      92456 & 0.41938 & 0.01520 & 0.49225 & 0.01435 & shallow_key_in=(168__60__1)_out=24_filters=4_short_shape=(3__1)_long_shape=(84__1)_dropout=0.3 \\
gtzan_key        & 5 &     203184 & 0.41579 & 0.03038 & 0.48713 & 0.02459 & shallow_key_in=(168__60__1)_out=24_filters=6_short_shape=(3__1)_long_shape=(84__1)_dropout=0.3 \\
gtzan_key        & 5 &     356920 & 0.40287 & 0.02235 & 0.48010 & 0.01929 & shallow_key_in=(168__60__1)_out=24_filters=8_short_shape=(3__1)_long_shape=(84__1)_dropout=0.1 \\
gtzan_key        & 5 &     793416 & 0.42297 & 0.01134 & 0.49711 & 0.01033 & shallow_key_in=(168__60__1)_out=24_filters=12_short_shape=(3__1)_long_shape=(84__1)_dropout=0.3 \\
}\gtzankeykeyshallowhalf

\pgfplotstableread[row sep=\\,col sep=&]{
Testset & Runs & Parameters & Acc & AccStd & Score & ScoreStd & Model \\
gtzan_key        & 5 &       5468 & 0.02799 & 0.00679 & 0.05938 & 0.00923 & shallow_tempo_in=(168__60__1)_out=24_filters=1_short_shape=(1__3)_long_shape=(1__60)_dropout=0.5 \\
gtzan_key        & 5 &      18592 & 0.03086 & 0.00988 & 0.06045 & 0.01303 & shallow_tempo_in=(168__60__1)_out=24_filters=2_short_shape=(1__3)_long_shape=(1__60)_dropout=0.5 \\
gtzan_key        & 5 &      67880 & 0.04856 & 0.00700 & 0.08426 & 0.00227 & shallow_tempo_in=(168__60__1)_out=24_filters=4_short_shape=(1__3)_long_shape=(1__60)_dropout=0.3 \\
gtzan_key        & 5 &     147888 & 0.03230 & 0.01029 & 0.06720 & 0.01819 & shallow_tempo_in=(168__60__1)_out=24_filters=6_short_shape=(1__3)_long_shape=(1__60)_dropout=0.3 \\
gtzan_key        & 5 &     258616 & 0.04163 & 0.01225 & 0.07447 & 0.01255 & shallow_tempo_in=(168__60__1)_out=24_filters=8_short_shape=(1__3)_long_shape=(1__60)_dropout=0.3 \\
gtzan_key        & 5 &     572232 & 0.02967 & 0.00607 & 0.06249 & 0.00513 & shallow_tempo_in=(168__60__1)_out=24_filters=12_short_shape=(1__3)_long_shape=(1__60)_dropout=0.1 \\
}\gtzankeytemposhallow

\pgfplotstableread[row sep=\\,col sep=&]{
Testset & Runs & Parameters & Acc & AccStd & Score & ScoreStd & Model \\
gtzan_key        & 5 &       3548 & 0.03971 & 0.01108 & 0.07911 & 0.01347 & shallow_tempo_in=(168__60__1)_out=24_filters=1_short_shape=(1__3)_long_shape=(1__30)_dropout=0.5 \\
gtzan_key        & 5 &      10912 & 0.03828 & 0.00951 & 0.07074 & 0.01661 & shallow_tempo_in=(168__60__1)_out=24_filters=2_short_shape=(1__3)_long_shape=(1__30)_dropout=0.5 \\
gtzan_key        & 5 &      37160 & 0.03517 & 0.01421 & 0.06632 & 0.01815 & shallow_tempo_in=(168__60__1)_out=24_filters=4_short_shape=(1__3)_long_shape=(1__30)_dropout=0.5 \\
gtzan_key        & 5 &      78768 & 0.03971 & 0.01027 & 0.08014 & 0.01605 & shallow_tempo_in=(168__60__1)_out=24_filters=6_short_shape=(1__3)_long_shape=(1__30)_dropout=0.5 \\
gtzan_key        & 5 &     135736 & 0.03541 & 0.01272 & 0.06981 & 0.02595 & shallow_tempo_in=(168__60__1)_out=24_filters=8_short_shape=(1__3)_long_shape=(1__30)_dropout=0.1 \\
gtzan_key        & 5 &     295752 & 0.04522 & 0.00735 & 0.08349 & 0.01553 & shallow_tempo_in=(168__60__1)_out=24_filters=12_short_shape=(1__3)_long_shape=(1__30)_dropout=0.3 \\
}\gtzankeytemposhallowhalf

\pgfplotstableread[row sep=\\,col sep=&]{
Testset & Runs & Parameters & Acc & AccStd & Score & ScoreStd & Model \\
gtzan_key        & 5 &       5378 & 0.50144 & 0.03740 & 0.56629 & 0.03325 & vgg_like_in=(168__60__1)_out=24_filters=2_pool_shape=(2__2)_max=True_filter_shapes=[(5__1)__(3__1)]_dropout=0.1 \\
gtzan_key        & 5 &      19572 & 0.50694 & 0.03108 & 0.57742 & 0.02646 & vgg_like_in=(168__60__1)_out=24_filters=4_pool_shape=(2__2)_max=True_filter_shapes=[(5__1)__(3__1)]_dropout=0.1 \\
gtzan_key        & 5 &      74480 & 0.48158 & 0.02034 & 0.56062 & 0.01691 & vgg_like_in=(168__60__1)_out=24_filters=8_pool_shape=(2__2)_max=True_filter_shapes=[(5__1)__(3__1)]_dropout=0.1 \\
gtzan_key        & 5 &     290376 & 0.44856 & 0.01498 & 0.53141 & 0.01442 & vgg_like_in=(168__60__1)_out=24_filters=16_pool_shape=(2__2)_max=True_filter_shapes=[(5__1)__(3__1)]_dropout=0.3 \\
gtzan_key        & 5 &     647712 & 0.44785 & 0.02004 & 0.53493 & 0.01848 & vgg_like_in=(168__60__1)_out=24_filters=24_pool_shape=(2__2)_max=True_filter_shapes=[(5__1)__(3__1)]_dropout=0.3 \\
}\gtzankeykeyvgg

\pgfplotstableread[row sep=\\,col sep=&]{
Testset & Runs & Parameters & Acc & AccStd & Score & ScoreStd & Model \\
gtzan_key        & 5 &       5378 & 0.11435 & 0.02399 & 0.15866 & 0.02514 & vgg_like_in=(168__60__1)_out=24_filters=2_pool_shape=(2__2)_max=True_filter_shapes=[(1__5)__(1__3)]_dropout=0.1 \\
gtzan_key        & 5 &      19572 & 0.23876 & 0.01281 & 0.30316 & 0.01064 & vgg_like_in=(168__60__1)_out=24_filters=4_pool_shape=(2__2)_max=True_filter_shapes=[(1__5)__(1__3)]_dropout=0.1 \\
gtzan_key        & 5 &      74480 & 0.34593 & 0.02602 & 0.40577 & 0.02716 & vgg_like_in=(168__60__1)_out=24_filters=8_pool_shape=(2__2)_max=True_filter_shapes=[(1__5)__(1__3)]_dropout=0.1 \\
gtzan_key        & 5 &     290376 & 0.34474 & 0.02491 & 0.41612 & 0.02775 & vgg_like_in=(168__60__1)_out=24_filters=16_pool_shape=(2__2)_max=True_filter_shapes=[(1__5)__(1__3)]_dropout=0.1 \\
gtzan_key        & 5 &     647712 & 0.38397 & 0.02427 & 0.45541 & 0.02339 & vgg_like_in=(168__60__1)_out=24_filters=24_pool_shape=(2__2)_max=True_filter_shapes=[(1__5)__(1__3)]_dropout=0.3 \\
}\gtzankeytempovgg

\pgfplotstableread[row sep=\\,col sep=&]{
Testset & Runs & Parameters & Acc & AccStd & Score & ScoreStd & Model \\
gtzan_key        & 5 &       5046 & 0.32823 & 0.05967 & 0.40529 & 0.05388 & vgg_like_in=(168__60__1)_out=24_filters=1_pool_shape=(2__2)_max=True_filter_shapes=[(5__5)__(3__3)]_dropout=0.1 \\
gtzan_key        & 5 &      19138 & 0.50837 & 0.05314 & 0.57411 & 0.04452 & vgg_like_in=(168__60__1)_out=24_filters=2_pool_shape=(2__2)_max=True_filter_shapes=[(5__5)__(3__3)]_dropout=0.3 \\
gtzan_key        & 5 &      74532 & 0.57129 & 0.03147 & 0.63069 & 0.02411 & vgg_like_in=(168__60__1)_out=24_filters=4_pool_shape=(2__2)_max=True_filter_shapes=[(5__5)__(3__3)]_dropout=0.3 \\
gtzan_key        & 5 &     294160 & 0.47775 & 0.02352 & 0.55591 & 0.01944 & vgg_like_in=(168__60__1)_out=24_filters=8_pool_shape=(2__2)_max=True_filter_shapes=[(5__5)__(3__3)]_dropout=0.1 \\
gtzan_key        & 5 &    1168776 & 0.54163 & 0.03289 & 0.61000 & 0.02790 & vgg_like_in=(168__60__1)_out=24_filters=16_pool_shape=(2__2)_max=True_filter_shapes=[(5__5)__(3__3)]_dropout=0.5 \\
gtzan_key        & 5 &    2623872 & 0.49856 & 0.02022 & 0.57127 & 0.01882 & vgg_like_in=(168__60__1)_out=24_filters=24_pool_shape=(2__2)_max=True_filter_shapes=[(5__5)__(3__3)]_dropout=0.5 \\
}\gtzankeysquarevgg


\pgfplotstableread[row sep=\\,col sep=&]{
Testset & Runs & Parameters & Acc1 & Std1 & Model \\
gs_new           & 5 &      19268 & 0.01928 & 0.00696 & shallow_key_in=(40__256__1)_out=256_filters=1_short_shape=(3__1)_long_shape=(40__1)_dropout=0.1 \\
gs_new           & 5 &      43400 & 0.07681 & 0.04765 & shallow_key_in=(40__256__1)_out=256_filters=2_short_shape=(3__1)_long_shape=(40__1)_dropout=0.3 \\
gs_new           & 5 &     107024 & 0.05602 & 0.02124 & shallow_key_in=(40__256__1)_out=256_filters=4_short_shape=(3__1)_long_shape=(40__1)_dropout=0.3 \\
gs_new           & 5 &     191128 & 0.04398 & 0.00973 & shallow_key_in=(40__256__1)_out=256_filters=6_short_shape=(3__1)_long_shape=(40__1)_dropout=0.3 \\
gs_new           & 5 &     295712 & 0.04970 & 0.01216 & shallow_key_in=(40__256__1)_out=256_filters=8_short_shape=(3__1)_long_shape=(40__1)_dropout=0.1 \\
gs_new           & 5 &     566320 & 0.04548 & 0.01855 & shallow_key_in=(40__256__1)_out=256_filters=12_short_shape=(3__1)_long_shape=(40__1)_dropout=0.5 \\
}\giantstepstempokeyshallow

\pgfplotstableread[row sep=\\,col sep=&]{
Testset & Runs & Parameters & Acc1 & Std1 & Model \\
gs_new           & 5 &      17988 & 0.03886 & 0.00525 & shallow_key_in=(40__256__1)_out=256_filters=1_short_shape=(3__1)_long_shape=(20__1)_dropout=0.3 \\
gs_new           & 5 &      38280 & 0.05542 & 0.02847 & shallow_key_in=(40__256__1)_out=256_filters=2_short_shape=(3__1)_long_shape=(20__1)_dropout=0.3 \\
gs_new           & 5 &      86544 & 0.09127 & 0.05320 & shallow_key_in=(40__256__1)_out=256_filters=4_short_shape=(3__1)_long_shape=(20__1)_dropout=0.3 \\
gs_new           & 5 &     145048 & 0.07922 & 0.03296 & shallow_key_in=(40__256__1)_out=256_filters=6_short_shape=(3__1)_long_shape=(20__1)_dropout=0.1 \\
gs_new           & 5 &     213792 & 0.05783 & 0.00717 & shallow_key_in=(40__256__1)_out=256_filters=8_short_shape=(3__1)_long_shape=(20__1)_dropout=0.5 \\
gs_new           & 5 &     382000 & 0.07139 & 0.02697 & shallow_key_in=(40__256__1)_out=256_filters=12_short_shape=(3__1)_long_shape=(20__1)_dropout=0.5 \\
}\giantstepstempokeyshallowhalf

\pgfplotstableread[row sep=\\,col sep=&]{
Testset & Runs & Parameters & Acc1 & Std1 & Model \\
gs_new           & 5 &      33092 & 0.54789 & 0.28002 & shallow_tempo_in=(40__256__1)_out=256_filters=1_short_shape=(1__3)_long_shape=(1__256)_dropout=0.1 \\
gs_new           & 5 &      98696 & 0.81777 & 0.04906 & shallow_tempo_in=(40__256__1)_out=256_filters=2_short_shape=(1__3)_long_shape=(1__256)_dropout=0.3 \\
gs_new           & 5 &     328208 & 0.87108 & 0.00584 & shallow_tempo_in=(40__256__1)_out=256_filters=4_short_shape=(1__3)_long_shape=(1__256)_dropout=0.5 \\
gs_new           & 5 &     688792 & 0.86205 & 0.01904 & shallow_tempo_in=(40__256__1)_out=256_filters=6_short_shape=(1__3)_long_shape=(1__256)_dropout=0.5 \\
gs_new           & 5 &    1180448 & 0.85934 & 0.01496 & shallow_tempo_in=(40__256__1)_out=256_filters=8_short_shape=(1__3)_long_shape=(1__256)_dropout=0.3 \\
gs_new           & 5 &    2556976 & 0.86506 & 0.01493 & shallow_tempo_in=(40__256__1)_out=256_filters=12_short_shape=(1__3)_long_shape=(1__256)_dropout=0.5 \\
}\giantstepstempotemposhallow

\pgfplotstableread[row sep=\\,col sep=&]{
Testset & Runs & Parameters & Acc1 & Std1 & Model \\
gs_new           & 5 &      24900 & 0.57139 & 0.27519 & shallow_tempo_in=(40__256__1)_out=256_filters=1_short_shape=(1__3)_long_shape=(1__128)_dropout=0.3 \\
gs_new           & 5 &      65928 & 0.82289 & 0.03241 & shallow_tempo_in=(40__256__1)_out=256_filters=2_short_shape=(1__3)_long_shape=(1__128)_dropout=0.5 \\
gs_new           & 5 &     197136 & 0.87892 & 0.00717 & shallow_tempo_in=(40__256__1)_out=256_filters=4_short_shape=(1__3)_long_shape=(1__128)_dropout=0.3 \\
gs_new           & 5 &     393880 & 0.86295 & 0.01917 & shallow_tempo_in=(40__256__1)_out=256_filters=6_short_shape=(1__3)_long_shape=(1__128)_dropout=0.5 \\
gs_new           & 5 &     656160 & 0.87982 & 0.01300 & shallow_tempo_in=(40__256__1)_out=256_filters=8_short_shape=(1__3)_long_shape=(1__128)_dropout=0.3 \\
gs_new           & 5 &    1377328 & 0.86988 & 0.01418 & shallow_tempo_in=(40__256__1)_out=256_filters=12_short_shape=(1__3)_long_shape=(1__128)_dropout=0.5 \\
}\giantstepstempotemposhallowhalf

\pgfplotstableread[row sep=\\,col sep=&]{
Testset & Runs & Parameters & Acc1 & Std1 & Model \\
gs_new           & 5 &       9322 & 0.63253 & 0.05314 & vgg_like_in=(40__256__1)_out=256_filters=2_pool_shape=(2__2)_max=True_filter_shapes=[(1__5)__(1__3)]_dropout=0.1 \\
gs_new           & 5 &      27228 & 0.77590 & 0.02844 & vgg_like_in=(40__256__1)_out=256_filters=4_pool_shape=(2__2)_max=True_filter_shapes=[(1__5)__(1__3)]_dropout=0.1 \\
gs_new           & 5 &      89560 & 0.85211 & 0.01778 & vgg_like_in=(40__256__1)_out=256_filters=8_pool_shape=(2__2)_max=True_filter_shapes=[(1__5)__(1__3)]_dropout=0.1 \\
gs_new           & 5 &     320304 & 0.88705 & 0.00610 & vgg_like_in=(40__256__1)_out=256_filters=16_pool_shape=(2__2)_max=True_filter_shapes=[(1__5)__(1__3)]_dropout=0.1 \\
gs_new           & 5 &     692488 & 0.88946 & 0.00388 & vgg_like_in=(40__256__1)_out=256_filters=24_pool_shape=(2__2)_max=True_filter_shapes=[(1__5)__(1__3)]_dropout=0.3 \\
}\giantstepstempotempovgg

\pgfplotstableread[row sep=\\,col sep=&]{
Testset & Runs & Parameters & Acc1 & Std1 & Model \\
gs_new           & 5 &       9322 & 0.23133 & 0.04260 & vgg_like_in=(40__256__1)_out=256_filters=2_pool_shape=(2__2)_max=True_filter_shapes=[(5__1)__(3__1)]_dropout=0.1 \\
gs_new           & 5 &      27228 & 0.30512 & 0.04020 & vgg_like_in=(40__256__1)_out=256_filters=4_pool_shape=(2__2)_max=True_filter_shapes=[(5__1)__(3__1)]_dropout=0.1 \\
gs_new           & 5 &      89560 & 0.41867 & 0.02837 & vgg_like_in=(40__256__1)_out=256_filters=8_pool_shape=(2__2)_max=True_filter_shapes=[(5__1)__(3__1)]_dropout=0.1 \\
gs_new           & 5 &     320304 & 0.46265 & 0.03601 & vgg_like_in=(40__256__1)_out=256_filters=16_pool_shape=(2__2)_max=True_filter_shapes=[(5__1)__(3__1)]_dropout=0.1 \\
gs_new           & 5 &     692488 & 0.49578 & 0.02549 & vgg_like_in=(40__256__1)_out=256_filters=24_pool_shape=(2__2)_max=True_filter_shapes=[(5__1)__(3__1)]_dropout=0.1 \\
}\giantstepstempokeyvgg

\pgfplotstableread[row sep=\\,col sep=&]{
Testset & Runs & Parameters & Acc1 & Std1 & Model \\
gs_new           & 5 &       7134 & 0.58102 & 0.02876 & vgg_like_in=(40__256__1)_out=256_filters=1_pool_shape=(2__2)_max=True_filter_shapes=[(5__5)__(3__3)]_dropout=0.1 \\
gs_new           & 5 &      23082 & 0.70753 & 0.01560 & vgg_like_in=(40__256__1)_out=256_filters=2_pool_shape=(2__2)_max=True_filter_shapes=[(5__5)__(3__3)]_dropout=0.1 \\
gs_new           & 5 &      82188 & 0.80361 & 0.03118 & vgg_like_in=(40__256__1)_out=256_filters=4_pool_shape=(2__2)_max=True_filter_shapes=[(5__5)__(3__3)]_dropout=0.1 \\
gs_new           & 5 &     309240 & 0.88163 & 0.01244 & vgg_like_in=(40__256__1)_out=256_filters=8_pool_shape=(2__2)_max=True_filter_shapes=[(5__5)__(3__3)]_dropout=0.3 \\
gs_new           & 5 &    1198704 & 0.88133 & 0.01328 & vgg_like_in=(40__256__1)_out=256_filters=16_pool_shape=(2__2)_max=True_filter_shapes=[(5__5)__(3__3)]_dropout=0.3 \\
gs_new           & 5 &    2668648 & 0.88584 & 0.00816 & vgg_like_in=(40__256__1)_out=256_filters=24_pool_shape=(2__2)_max=True_filter_shapes=[(5__5)__(3__3)]_dropout=0.5 \\
}\giantstepstemposquarevgg


\pgfplotstableread[row sep=\\,col sep=&]{
Testset & Runs & Parameters & Acc & AccStd & Score & ScoreStd & Model \\
giantsteps-key   & 5 &      12380 & 0.43046 & 0.03615 & 0.50778 & 0.03272 & shallow_key_in=(168__60__1)_out=24_filters=1_short_shape=(3__1)_long_shape=(168__1)_dropout=0.1 \\
giantsteps-key   & 5 &      46240 & 0.48940 & 0.02453 & 0.56096 & 0.01993 & shallow_key_in=(168__60__1)_out=24_filters=2_short_shape=(3__1)_long_shape=(168__1)_dropout=0.3 \\
giantsteps-key   & 5 &     178472 & 0.50828 & 0.03756 & 0.57894 & 0.02898 & shallow_key_in=(168__60__1)_out=24_filters=4_short_shape=(3__1)_long_shape=(168__1)_dropout=0.5 \\
giantsteps-key   & 5 &     396720 & 0.50033 & 0.06316 & 0.57252 & 0.05410 & shallow_key_in=(168__60__1)_out=24_filters=6_short_shape=(3__1)_long_shape=(168__1)_dropout=0.3 \\
giantsteps-key   & 5 &     700984 & 0.53444 & 0.03975 & 0.60043 & 0.03152 & shallow_key_in=(168__60__1)_out=24_filters=8_short_shape=(3__1)_long_shape=(168__1)_dropout=0.5 \\
giantsteps-key   & 5 &    1567560 & 0.47384 & 0.04090 & 0.54917 & 0.03444 & shallow_key_in=(168__60__1)_out=24_filters=12_short_shape=(3__1)_long_shape=(168__1)_dropout=0.3 \\
}\giantstepskeykeyshallow

\pgfplotstableread[row sep=\\,col sep=&]{
Testset & Runs & Parameters & Acc & AccStd & Score & ScoreStd & Model \\
giantsteps-key   & 5 &       7004 & 0.30166 & 0.03757 & 0.39238 & 0.03534 & shallow_key_in=(168__60__1)_out=24_filters=1_short_shape=(3__1)_long_shape=(84__1)_dropout=0.1 \\
giantsteps-key   & 5 &      24736 & 0.47616 & 0.02861 & 0.54258 & 0.02455 & shallow_key_in=(168__60__1)_out=24_filters=2_short_shape=(3__1)_long_shape=(84__1)_dropout=0.3 \\
giantsteps-key   & 5 &      92456 & 0.47649 & 0.02664 & 0.54493 & 0.02166 & shallow_key_in=(168__60__1)_out=24_filters=4_short_shape=(3__1)_long_shape=(84__1)_dropout=0.3 \\
giantsteps-key   & 5 &     203184 & 0.47649 & 0.06320 & 0.54371 & 0.05365 & shallow_key_in=(168__60__1)_out=24_filters=6_short_shape=(3__1)_long_shape=(84__1)_dropout=0.3 \\
giantsteps-key   & 5 &     356920 & 0.40099 & 0.05853 & 0.48070 & 0.04916 & shallow_key_in=(168__60__1)_out=24_filters=8_short_shape=(3__1)_long_shape=(84__1)_dropout=0.1 \\
giantsteps-key   & 5 &     793416 & 0.40993 & 0.07210 & 0.48725 & 0.06195 & shallow_key_in=(168__60__1)_out=24_filters=12_short_shape=(3__1)_long_shape=(84__1)_dropout=0.3 \\
}\giantstepskeykeyshallowhalf

\pgfplotstableread[row sep=\\,col sep=&]{
Testset & Runs & Parameters & Acc & AccStd & Score & ScoreStd & Model \\
giantsteps-key   & 5 &       5468 & 0.01026 & 0.00354 & 0.04414 & 0.00655 & shallow_tempo_in=(168__60__1)_out=24_filters=1_short_shape=(1__3)_long_shape=(1__60)_dropout=0.5 \\
giantsteps-key   & 5 &      18592 & 0.01192 & 0.00528 & 0.04023 & 0.01328 & shallow_tempo_in=(168__60__1)_out=24_filters=2_short_shape=(1__3)_long_shape=(1__60)_dropout=0.5 \\
giantsteps-key   & 5 &      67880 & 0.01689 & 0.00354 & 0.04904 & 0.00227 & shallow_tempo_in=(168__60__1)_out=24_filters=4_short_shape=(1__3)_long_shape=(1__60)_dropout=0.3 \\
giantsteps-key   & 5 &     147888 & 0.02219 & 0.01071 & 0.05639 & 0.01091 & shallow_tempo_in=(168__60__1)_out=24_filters=6_short_shape=(1__3)_long_shape=(1__60)_dropout=0.3 \\
giantsteps-key   & 5 &     258616 & 0.03477 & 0.01068 & 0.07811 & 0.01631 & shallow_tempo_in=(168__60__1)_out=24_filters=8_short_shape=(1__3)_long_shape=(1__60)_dropout=0.3 \\
giantsteps-key   & 5 &     572232 & 0.01391 & 0.00169 & 0.04772 & 0.00877 & shallow_tempo_in=(168__60__1)_out=24_filters=12_short_shape=(1__3)_long_shape=(1__60)_dropout=0.1 \\
}\giantstepskeytemposhallow

\pgfplotstableread[row sep=\\,col sep=&]{
Testset & Runs & Parameters & Acc & AccStd & Score & ScoreStd & Model \\
giantsteps-key   & 5 &       3548 & 0.01391 & 0.00427 & 0.05371 & 0.00885 & shallow_tempo_in=(168__60__1)_out=24_filters=1_short_shape=(1__3)_long_shape=(1__30)_dropout=0.5 \\
giantsteps-key   & 5 &      10912 & 0.01722 & 0.00867 & 0.03990 & 0.01283 & shallow_tempo_in=(168__60__1)_out=24_filters=2_short_shape=(1__3)_long_shape=(1__30)_dropout=0.5 \\
giantsteps-key   & 5 &      37160 & 0.02517 & 0.01209 & 0.05427 & 0.01135 & shallow_tempo_in=(168__60__1)_out=24_filters=4_short_shape=(1__3)_long_shape=(1__30)_dropout=0.5 \\
giantsteps-key   & 5 &      78768 & 0.02815 & 0.01230 & 0.06056 & 0.01708 & shallow_tempo_in=(168__60__1)_out=24_filters=6_short_shape=(1__3)_long_shape=(1__30)_dropout=0.5 \\
giantsteps-key   & 5 &     135736 & 0.01358 & 0.00587 & 0.04934 & 0.01336 & shallow_tempo_in=(168__60__1)_out=24_filters=8_short_shape=(1__3)_long_shape=(1__30)_dropout=0.1 \\
giantsteps-key   & 5 &     295752 & 0.02086 & 0.00828 & 0.05818 & 0.01080 & shallow_tempo_in=(168__60__1)_out=24_filters=12_short_shape=(1__3)_long_shape=(1__30)_dropout=0.3 \\
}\giantstepskeytemposhallowhalf

\pgfplotstableread[row sep=\\,col sep=&]{
Testset & Runs & Parameters & Acc & AccStd & Score & ScoreStd & Model \\
giantsteps-key   & 5 &       5378 & 0.50695 & 0.02492 & 0.57599 & 0.01879 & vgg_like_in=(168__60__1)_out=24_filters=2_pool_shape=(2__2)_max=True_filter_shapes=[(5__1)__(3__1)]_dropout=0.1 \\
giantsteps-key   & 5 &      19572 & 0.55199 & 0.02281 & 0.61325 & 0.01820 & vgg_like_in=(168__60__1)_out=24_filters=4_pool_shape=(2__2)_max=True_filter_shapes=[(5__1)__(3__1)]_dropout=0.1 \\
giantsteps-key   & 5 &      74480 & 0.58907 & 0.01403 & 0.64586 & 0.01104 & vgg_like_in=(168__60__1)_out=24_filters=8_pool_shape=(2__2)_max=True_filter_shapes=[(5__1)__(3__1)]_dropout=0.1 \\
giantsteps-key   & 5 &     290376 & 0.55033 & 0.02479 & 0.61175 & 0.02097 & vgg_like_in=(168__60__1)_out=24_filters=16_pool_shape=(2__2)_max=True_filter_shapes=[(5__1)__(3__1)]_dropout=0.3 \\
giantsteps-key   & 5 &     647712 & 0.55397 & 0.02671 & 0.61745 & 0.02073 & vgg_like_in=(168__60__1)_out=24_filters=24_pool_shape=(2__2)_max=True_filter_shapes=[(5__1)__(3__1)]_dropout=0.3 \\
}\giantstepskeykeyvgg

\pgfplotstableread[row sep=\\,col sep=&]{
Testset & Runs & Parameters & Acc & AccStd & Score & ScoreStd & Model \\
giantsteps-key   & 5 &       5378 & 0.09967 & 0.01441 & 0.14149 & 0.01677 & vgg_like_in=(168__60__1)_out=24_filters=2_pool_shape=(2__2)_max=True_filter_shapes=[(1__5)__(1__3)]_dropout=0.1 \\
giantsteps-key   & 5 &      19572 & 0.15662 & 0.03057 & 0.22255 & 0.02920 & vgg_like_in=(168__60__1)_out=24_filters=4_pool_shape=(2__2)_max=True_filter_shapes=[(1__5)__(1__3)]_dropout=0.1 \\
giantsteps-key   & 5 &      74480 & 0.32417 & 0.05504 & 0.39046 & 0.04648 & vgg_like_in=(168__60__1)_out=24_filters=8_pool_shape=(2__2)_max=True_filter_shapes=[(1__5)__(1__3)]_dropout=0.1 \\
giantsteps-key   & 5 &     290376 & 0.36093 & 0.09337 & 0.43636 & 0.08137 & vgg_like_in=(168__60__1)_out=24_filters=16_pool_shape=(2__2)_max=True_filter_shapes=[(1__5)__(1__3)]_dropout=0.1 \\
giantsteps-key   & 5 &     647712 & 0.46788 & 0.04282 & 0.53093 & 0.03710 & vgg_like_in=(168__60__1)_out=24_filters=24_pool_shape=(2__2)_max=True_filter_shapes=[(1__5)__(1__3)]_dropout=0.3 \\
}\giantstepskeytempovgg

\pgfplotstableread[row sep=\\,col sep=&]{
Testset & Runs & Parameters & Acc & AccStd & Score & ScoreStd & Model \\
giantsteps-key   & 5 &       5046 & 0.20894 & 0.06084 & 0.29460 & 0.05583 & vgg_like_in=(168__60__1)_out=24_filters=1_pool_shape=(2__2)_max=True_filter_shapes=[(5__5)__(3__3)]_dropout=0.1 \\
giantsteps-key   & 5 &      19138 & 0.47318 & 0.07149 & 0.54507 & 0.05909 & vgg_like_in=(168__60__1)_out=24_filters=2_pool_shape=(2__2)_max=True_filter_shapes=[(5__5)__(3__3)]_dropout=0.3 \\
giantsteps-key   & 5 &      74532 & 0.60596 & 0.02164 & 0.65748 & 0.01745 & vgg_like_in=(168__60__1)_out=24_filters=4_pool_shape=(2__2)_max=True_filter_shapes=[(5__5)__(3__3)]_dropout=0.3 \\
giantsteps-key   & 5 &     294160 & 0.58046 & 0.01442 & 0.63970 & 0.01125 & vgg_like_in=(168__60__1)_out=24_filters=8_pool_shape=(2__2)_max=True_filter_shapes=[(5__5)__(3__3)]_dropout=0.1 \\
giantsteps-key   & 5 &    1168776 & 0.62450 & 0.02581 & 0.67510 & 0.02151 & vgg_like_in=(168__60__1)_out=24_filters=16_pool_shape=(2__2)_max=True_filter_shapes=[(5__5)__(3__3)]_dropout=0.5 \\
giantsteps-key   & 5 &    2623872 & 0.58510 & 0.03930 & 0.64450 & 0.03226 & vgg_like_in=(168__60__1)_out=24_filters=24_pool_shape=(2__2)_max=True_filter_shapes=[(5__5)__(3__3)]_dropout=0.5 \\
}\giantstepskeysquarevgg


\pgfplotstableread[row sep=\\,col sep=&]{
Testset & Runs & Parameters & Acc1 & Std1 & Model \\
lmd_tempo_test   & 5 &      19268 & 0.06925 & 0.01299 & shallow_key_in=(40__256__1)_out=256_filters=1_short_shape=(3__1)_long_shape=(40__1)_dropout=0.1 \\
lmd_tempo_test   & 5 &      43400 & 0.10693 & 0.06564 & shallow_key_in=(40__256__1)_out=256_filters=2_short_shape=(3__1)_long_shape=(40__1)_dropout=0.3 \\
lmd_tempo_test   & 5 &     107024 & 0.11191 & 0.02178 & shallow_key_in=(40__256__1)_out=256_filters=4_short_shape=(3__1)_long_shape=(40__1)_dropout=0.3 \\
lmd_tempo_test   & 5 &     191128 & 0.09030 & 0.02092 & shallow_key_in=(40__256__1)_out=256_filters=6_short_shape=(3__1)_long_shape=(40__1)_dropout=0.3 \\
lmd_tempo_test   & 5 &     295712 & 0.09529 & 0.01102 & shallow_key_in=(40__256__1)_out=256_filters=8_short_shape=(3__1)_long_shape=(40__1)_dropout=0.1 \\
lmd_tempo_test   & 5 &     566320 & 0.09418 & 0.02102 & shallow_key_in=(40__256__1)_out=256_filters=12_short_shape=(3__1)_long_shape=(40__1)_dropout=0.5 \\
}\lmdtempokeyshallow

\pgfplotstableread[row sep=\\,col sep=&]{
Testset & Runs & Parameters & Acc1 & Std1 & Model \\
lmd_tempo_test   & 5 &      17988 & 0.07867 & 0.02608 & shallow_key_in=(40__256__1)_out=256_filters=1_short_shape=(3__1)_long_shape=(20__1)_dropout=0.3 \\
lmd_tempo_test   & 5 &      38280 & 0.10028 & 0.03876 & shallow_key_in=(40__256__1)_out=256_filters=2_short_shape=(3__1)_long_shape=(20__1)_dropout=0.3 \\
lmd_tempo_test   & 5 &      86544 & 0.11801 & 0.03467 & shallow_key_in=(40__256__1)_out=256_filters=4_short_shape=(3__1)_long_shape=(20__1)_dropout=0.3 \\
lmd_tempo_test   & 5 &     145048 & 0.09806 & 0.02649 & shallow_key_in=(40__256__1)_out=256_filters=6_short_shape=(3__1)_long_shape=(20__1)_dropout=0.1 \\
lmd_tempo_test   & 5 &     213792 & 0.10360 & 0.03593 & shallow_key_in=(40__256__1)_out=256_filters=8_short_shape=(3__1)_long_shape=(20__1)_dropout=0.5 \\
lmd_tempo_test   & 5 &     382000 & 0.12687 & 0.05201 & shallow_key_in=(40__256__1)_out=256_filters=12_short_shape=(3__1)_long_shape=(20__1)_dropout=0.5 \\
}\lmdtempokeyshallowhalf

\pgfplotstableread[row sep=\\,col sep=&]{
Testset & Runs & Parameters & Acc1 & Std1 & Model \\
lmd_tempo_test   & 5 &      33092 & 0.75346 & 0.35331 & shallow_tempo_in=(40__256__1)_out=256_filters=1_short_shape=(1__3)_long_shape=(1__256)_dropout=0.1 \\
lmd_tempo_test   & 5 &      98696 & 0.90416 & 0.01859 & shallow_tempo_in=(40__256__1)_out=256_filters=2_short_shape=(1__3)_long_shape=(1__256)_dropout=0.3 \\
lmd_tempo_test   & 5 &     328208 & 0.90970 & 0.02302 & shallow_tempo_in=(40__256__1)_out=256_filters=4_short_shape=(1__3)_long_shape=(1__256)_dropout=0.5 \\
lmd_tempo_test   & 5 &     688792 & 0.92576 & 0.01304 & shallow_tempo_in=(40__256__1)_out=256_filters=6_short_shape=(1__3)_long_shape=(1__256)_dropout=0.5 \\
lmd_tempo_test   & 5 &    1180448 & 0.91690 & 0.01527 & shallow_tempo_in=(40__256__1)_out=256_filters=8_short_shape=(1__3)_long_shape=(1__256)_dropout=0.3 \\
lmd_tempo_test   & 5 &    2556976 & 0.94017 & 0.00953 & shallow_tempo_in=(40__256__1)_out=256_filters=12_short_shape=(1__3)_long_shape=(1__256)_dropout=0.5 \\
}\lmdtempotemposhallow

\pgfplotstableread[row sep=\\,col sep=&]{
Testset & Runs & Parameters & Acc1 & Std1 & Model \\
lmd_tempo_test   & 5 &      24900 & 0.75845 & 0.32947 & shallow_tempo_in=(40__256__1)_out=256_filters=1_short_shape=(1__3)_long_shape=(1__128)_dropout=0.3 \\
lmd_tempo_test   & 5 &      65928 & 0.94183 & 0.00392 & shallow_tempo_in=(40__256__1)_out=256_filters=2_short_shape=(1__3)_long_shape=(1__128)_dropout=0.5 \\
lmd_tempo_test   & 5 &     197136 & 0.93629 & 0.00893 & shallow_tempo_in=(40__256__1)_out=256_filters=4_short_shape=(1__3)_long_shape=(1__128)_dropout=0.3 \\
lmd_tempo_test   & 5 &     393880 & 0.92964 & 0.01826 & shallow_tempo_in=(40__256__1)_out=256_filters=6_short_shape=(1__3)_long_shape=(1__128)_dropout=0.5 \\
lmd_tempo_test   & 5 &     656160 & 0.93130 & 0.01219 & shallow_tempo_in=(40__256__1)_out=256_filters=8_short_shape=(1__3)_long_shape=(1__128)_dropout=0.3 \\
lmd_tempo_test   & 5 &    1377328 & 0.93906 & 0.01476 & shallow_tempo_in=(40__256__1)_out=256_filters=12_short_shape=(1__3)_long_shape=(1__128)_dropout=0.5 \\
}\lmdtempotemposhallowhalf

\pgfplotstableread[row sep=\\,col sep=&]{
Testset & Runs & Parameters & Acc1 & Std1 & Model \\
lmd_tempo_test   & 5 &       9322 & 0.88643 & 0.02657 & vgg_like_in=(40__256__1)_out=256_filters=2_pool_shape=(2__2)_max=True_filter_shapes=[(1__5)__(1__3)]_dropout=0.1 \\
lmd_tempo_test   & 5 &      27228 & 0.93795 & 0.01375 & vgg_like_in=(40__256__1)_out=256_filters=4_pool_shape=(2__2)_max=True_filter_shapes=[(1__5)__(1__3)]_dropout=0.1 \\
lmd_tempo_test   & 5 &      89560 & 0.93795 & 0.01553 & vgg_like_in=(40__256__1)_out=256_filters=8_pool_shape=(2__2)_max=True_filter_shapes=[(1__5)__(1__3)]_dropout=0.1 \\
lmd_tempo_test   & 5 &     320304 & 0.94515 & 0.00709 & vgg_like_in=(40__256__1)_out=256_filters=16_pool_shape=(2__2)_max=True_filter_shapes=[(1__5)__(1__3)]_dropout=0.1 \\
lmd_tempo_test   & 5 &     692488 & 0.95125 & 0.01503 & vgg_like_in=(40__256__1)_out=256_filters=24_pool_shape=(2__2)_max=True_filter_shapes=[(1__5)__(1__3)]_dropout=0.3 \\
}\lmdtempotempovgg

\pgfplotstableread[row sep=\\,col sep=&]{
Testset & Runs & Parameters & Acc1 & Std1 & Model \\
lmd_tempo_test   & 5 &       9322 & 0.26870 & 0.04316 & vgg_like_in=(40__256__1)_out=256_filters=2_pool_shape=(2__2)_max=True_filter_shapes=[(5__1)__(3__1)]_dropout=0.1 \\
lmd_tempo_test   & 5 &      27228 & 0.40166 & 0.01903 & vgg_like_in=(40__256__1)_out=256_filters=4_pool_shape=(2__2)_max=True_filter_shapes=[(5__1)__(3__1)]_dropout=0.1 \\
lmd_tempo_test   & 5 &      89560 & 0.60942 & 0.00581 & vgg_like_in=(40__256__1)_out=256_filters=8_pool_shape=(2__2)_max=True_filter_shapes=[(5__1)__(3__1)]_dropout=0.1 \\
lmd_tempo_test   & 5 &     320304 & 0.69917 & 0.02751 & vgg_like_in=(40__256__1)_out=256_filters=16_pool_shape=(2__2)_max=True_filter_shapes=[(5__1)__(3__1)]_dropout=0.1 \\
lmd_tempo_test   & 5 &     692488 & 0.73019 & 0.02393 & vgg_like_in=(40__256__1)_out=256_filters=24_pool_shape=(2__2)_max=True_filter_shapes=[(5__1)__(3__1)]_dropout=0.1 \\
}\lmdtempokeyvgg

\pgfplotstableread[row sep=\\,col sep=&]{
Testset & Runs & Parameters & Acc1 & Std1 & Model \\
lmd_tempo_test   & 5 &       7134 & 0.89086 & 0.01341 & vgg_like_in=(40__256__1)_out=256_filters=1_pool_shape=(2__2)_max=True_filter_shapes=[(5__5)__(3__3)]_dropout=0.1 \\
lmd_tempo_test   & 5 &      23082 & 0.93573 & 0.00688 & vgg_like_in=(40__256__1)_out=256_filters=2_pool_shape=(2__2)_max=True_filter_shapes=[(5__5)__(3__3)]_dropout=0.1 \\
lmd_tempo_test   & 5 &      82188 & 0.95734 & 0.01130 & vgg_like_in=(40__256__1)_out=256_filters=4_pool_shape=(2__2)_max=True_filter_shapes=[(5__5)__(3__3)]_dropout=0.1 \\
lmd_tempo_test   & 5 &     309240 & 0.95956 & 0.00543 & vgg_like_in=(40__256__1)_out=256_filters=8_pool_shape=(2__2)_max=True_filter_shapes=[(5__5)__(3__3)]_dropout=0.3 \\
lmd_tempo_test   & 5 &    1198704 & 0.96177 & 0.00443 & vgg_like_in=(40__256__1)_out=256_filters=16_pool_shape=(2__2)_max=True_filter_shapes=[(5__5)__(3__3)]_dropout=0.3 \\
lmd_tempo_test   & 5 &    2668648 & 0.95956 & 0.01196 & vgg_like_in=(40__256__1)_out=256_filters=24_pool_shape=(2__2)_max=True_filter_shapes=[(5__5)__(3__3)]_dropout=0.5 \\
}\lmdtemposquarevgg


\pgfplotstableread[row sep=\\,col sep=&]{
Testset & Runs & Parameters & Acc & AccStd & Score & ScoreStd & Model \\
lmd_key_test     & 5 &      12380 & 0.66132 & 0.01913 & 0.70229 & 0.01780 & shallow_key_in=(168__60__1)_out=24_filters=1_short_shape=(3__1)_long_shape=(168__1)_dropout=0.1 \\
lmd_key_test     & 5 &      46240 & 0.66648 & 0.00776 & 0.70662 & 0.00731 & shallow_key_in=(168__60__1)_out=24_filters=2_short_shape=(3__1)_long_shape=(168__1)_dropout=0.3 \\
lmd_key_test     & 5 &     178472 & 0.67135 & 0.00895 & 0.71080 & 0.00759 & shallow_key_in=(168__60__1)_out=24_filters=4_short_shape=(3__1)_long_shape=(168__1)_dropout=0.5 \\
lmd_key_test     & 5 &     396720 & 0.67765 & 0.00471 & 0.71504 & 0.00285 & shallow_key_in=(168__60__1)_out=24_filters=6_short_shape=(3__1)_long_shape=(168__1)_dropout=0.3 \\
lmd_key_test     & 5 &     700984 & 0.67822 & 0.00619 & 0.71860 & 0.00678 & shallow_key_in=(168__60__1)_out=24_filters=8_short_shape=(3__1)_long_shape=(168__1)_dropout=0.5 \\
lmd_key_test     & 5 &    1567560 & 0.68367 & 0.00852 & 0.72106 & 0.00676 & shallow_key_in=(168__60__1)_out=24_filters=12_short_shape=(3__1)_long_shape=(168__1)_dropout=0.3 \\
}\lmdkeykeyshallow

\pgfplotstableread[row sep=\\,col sep=&]{
Testset & Runs & Parameters & Acc & AccStd & Score & ScoreStd & Model \\
lmd_key_test     & 5 &       7004 & 0.63582 & 0.01840 & 0.67771 & 0.01537 & shallow_key_in=(168__60__1)_out=24_filters=1_short_shape=(3__1)_long_shape=(84__1)_dropout=0.1 \\
lmd_key_test     & 5 &      24736 & 0.63438 & 0.00508 & 0.67997 & 0.00517 & shallow_key_in=(168__60__1)_out=24_filters=2_short_shape=(3__1)_long_shape=(84__1)_dropout=0.3 \\
lmd_key_test     & 5 &      92456 & 0.64613 & 0.01780 & 0.68900 & 0.01717 & shallow_key_in=(168__60__1)_out=24_filters=4_short_shape=(3__1)_long_shape=(84__1)_dropout=0.3 \\
lmd_key_test     & 5 &     203184 & 0.65072 & 0.02320 & 0.69261 & 0.02046 & shallow_key_in=(168__60__1)_out=24_filters=6_short_shape=(3__1)_long_shape=(84__1)_dropout=0.3 \\
lmd_key_test     & 5 &     356920 & 0.66160 & 0.01831 & 0.70195 & 0.01808 & shallow_key_in=(168__60__1)_out=24_filters=8_short_shape=(3__1)_long_shape=(84__1)_dropout=0.1 \\
lmd_key_test     & 5 &     793416 & 0.66734 & 0.00866 & 0.70473 & 0.00948 & shallow_key_in=(168__60__1)_out=24_filters=12_short_shape=(3__1)_long_shape=(84__1)_dropout=0.3 \\
}\lmdkeykeyshallowhalf

\pgfplotstableread[row sep=\\,col sep=&]{
Testset & Runs & Parameters & Acc & AccStd & Score & ScoreStd & Model \\
lmd_key_test     & 5 &       5468 & 0.08252 & 0.03066 & 0.12057 & 0.03979 & shallow_tempo_in=(168__60__1)_out=24_filters=1_short_shape=(1__3)_long_shape=(1__60)_dropout=0.5 \\
lmd_key_test     & 5 &      18592 & 0.07650 & 0.02209 & 0.10653 & 0.02105 & shallow_tempo_in=(168__60__1)_out=24_filters=2_short_shape=(1__3)_long_shape=(1__60)_dropout=0.5 \\
lmd_key_test     & 5 &      67880 & 0.11003 & 0.03661 & 0.14269 & 0.03514 & shallow_tempo_in=(168__60__1)_out=24_filters=4_short_shape=(1__3)_long_shape=(1__60)_dropout=0.3 \\
lmd_key_test     & 5 &     147888 & 0.07708 & 0.04214 & 0.10865 & 0.03343 & shallow_tempo_in=(168__60__1)_out=24_filters=6_short_shape=(1__3)_long_shape=(1__60)_dropout=0.3 \\
lmd_key_test     & 5 &     258616 & 0.10860 & 0.02155 & 0.13951 & 0.02746 & shallow_tempo_in=(168__60__1)_out=24_filters=8_short_shape=(1__3)_long_shape=(1__60)_dropout=0.3 \\
lmd_key_test     & 5 &     572232 & 0.09054 & 0.04099 & 0.12725 & 0.05116 & shallow_tempo_in=(168__60__1)_out=24_filters=12_short_shape=(1__3)_long_shape=(1__60)_dropout=0.1 \\
}\lmdkeytemposhallow

\pgfplotstableread[row sep=\\,col sep=&]{
Testset & Runs & Parameters & Acc & AccStd & Score & ScoreStd & Model \\
lmd_key_test     & 5 &       3548 & 0.10688 & 0.02728 & 0.14003 & 0.00996 & shallow_tempo_in=(168__60__1)_out=24_filters=1_short_shape=(1__3)_long_shape=(1__30)_dropout=0.5 \\
lmd_key_test     & 5 &      10912 & 0.09083 & 0.05765 & 0.13407 & 0.05416 & shallow_tempo_in=(168__60__1)_out=24_filters=2_short_shape=(1__3)_long_shape=(1__30)_dropout=0.5 \\
lmd_key_test     & 5 &      37160 & 0.10029 & 0.04112 & 0.13596 & 0.04066 & shallow_tempo_in=(168__60__1)_out=24_filters=4_short_shape=(1__3)_long_shape=(1__30)_dropout=0.5 \\
lmd_key_test     & 5 &      78768 & 0.10630 & 0.04030 & 0.14702 & 0.04244 & shallow_tempo_in=(168__60__1)_out=24_filters=6_short_shape=(1__3)_long_shape=(1__30)_dropout=0.5 \\
lmd_key_test     & 5 &     135736 & 0.07364 & 0.04809 & 0.09968 & 0.03268 & shallow_tempo_in=(168__60__1)_out=24_filters=8_short_shape=(1__3)_long_shape=(1__30)_dropout=0.1 \\
lmd_key_test     & 5 &     295752 & 0.10831 & 0.02257 & 0.14287 & 0.03379 & shallow_tempo_in=(168__60__1)_out=24_filters=12_short_shape=(1__3)_long_shape=(1__30)_dropout=0.3 \\
}\lmdkeytemposhallowhalf

\pgfplotstableread[row sep=\\,col sep=&]{
Testset & Runs & Parameters & Acc & AccStd & Score & ScoreStd & Model \\
lmd_key_test     & 5 &       5378 & 0.68968 & 0.00957 & 0.72779 & 0.00830 & vgg_like_in=(168__60__1)_out=24_filters=2_pool_shape=(2__2)_max=True_filter_shapes=[(5__1)__(3__1)]_dropout=0.1 \\
lmd_key_test     & 5 &      19572 & 0.69484 & 0.00731 & 0.73103 & 0.00523 & vgg_like_in=(168__60__1)_out=24_filters=4_pool_shape=(2__2)_max=True_filter_shapes=[(5__1)__(3__1)]_dropout=0.1 \\
lmd_key_test     & 5 &      74480 & 0.70401 & 0.00517 & 0.73779 & 0.00430 & vgg_like_in=(168__60__1)_out=24_filters=8_pool_shape=(2__2)_max=True_filter_shapes=[(5__1)__(3__1)]_dropout=0.1 \\
lmd_key_test     & 5 &     290376 & 0.70315 & 0.00948 & 0.73754 & 0.00727 & vgg_like_in=(168__60__1)_out=24_filters=16_pool_shape=(2__2)_max=True_filter_shapes=[(5__1)__(3__1)]_dropout=0.3 \\
lmd_key_test     & 5 &     647712 & 0.71261 & 0.00214 & 0.74487 & 0.00208 & vgg_like_in=(168__60__1)_out=24_filters=24_pool_shape=(2__2)_max=True_filter_shapes=[(5__1)__(3__1)]_dropout=0.3 \\
}\lmdkeykeyvgg

\pgfplotstableread[row sep=\\,col sep=&]{
Testset & Runs & Parameters & Acc & AccStd & Score & ScoreStd & Model \\
lmd_key_test     & 5 &       5378 & 0.16332 & 0.05867 & 0.20393 & 0.06086 & vgg_like_in=(168__60__1)_out=24_filters=2_pool_shape=(2__2)_max=True_filter_shapes=[(1__5)__(1__3)]_dropout=0.1 \\
lmd_key_test     & 5 &      19572 & 0.38997 & 0.06439 & 0.44252 & 0.06069 & vgg_like_in=(168__60__1)_out=24_filters=4_pool_shape=(2__2)_max=True_filter_shapes=[(1__5)__(1__3)]_dropout=0.1 \\
lmd_key_test     & 5 &      74480 & 0.53639 & 0.00717 & 0.58682 & 0.01090 & vgg_like_in=(168__60__1)_out=24_filters=8_pool_shape=(2__2)_max=True_filter_shapes=[(1__5)__(1__3)]_dropout=0.1 \\
lmd_key_test     & 5 &     290376 & 0.59570 & 0.02073 & 0.64232 & 0.01678 & vgg_like_in=(168__60__1)_out=24_filters=16_pool_shape=(2__2)_max=True_filter_shapes=[(1__5)__(1__3)]_dropout=0.1 \\
lmd_key_test     & 5 &     647712 & 0.60659 & 0.00431 & 0.65026 & 0.00487 & vgg_like_in=(168__60__1)_out=24_filters=24_pool_shape=(2__2)_max=True_filter_shapes=[(1__5)__(1__3)]_dropout=0.3 \\
}\lmdkeytempovgg

\pgfplotstableread[row sep=\\,col sep=&]{
Testset & Runs & Parameters & Acc & AccStd & Score & ScoreStd & Model \\
lmd_key_test     & 5 &       5046 & 0.47163 & 0.07070 & 0.52252 & 0.06024 & vgg_like_in=(168__60__1)_out=24_filters=1_pool_shape=(2__2)_max=True_filter_shapes=[(5__5)__(3__3)]_dropout=0.1 \\
lmd_key_test     & 5 &      19138 & 0.65731 & 0.01295 & 0.69905 & 0.01173 & vgg_like_in=(168__60__1)_out=24_filters=2_pool_shape=(2__2)_max=True_filter_shapes=[(5__5)__(3__3)]_dropout=0.3 \\
lmd_key_test     & 5 &      74532 & 0.67307 & 0.03266 & 0.71513 & 0.02546 & vgg_like_in=(168__60__1)_out=24_filters=4_pool_shape=(2__2)_max=True_filter_shapes=[(5__5)__(3__3)]_dropout=0.3 \\
lmd_key_test     & 5 &     294160 & 0.70315 & 0.01321 & 0.73771 & 0.01141 & vgg_like_in=(168__60__1)_out=24_filters=8_pool_shape=(2__2)_max=True_filter_shapes=[(5__5)__(3__3)]_dropout=0.1 \\
lmd_key_test     & 5 &    1168776 & 0.68739 & 0.01573 & 0.72728 & 0.01318 & vgg_like_in=(168__60__1)_out=24_filters=16_pool_shape=(2__2)_max=True_filter_shapes=[(5__5)__(3__3)]_dropout=0.5 \\
lmd_key_test     & 5 &    2623872 & 0.68911 & 0.02547 & 0.72668 & 0.02278 & vgg_like_in=(168__60__1)_out=24_filters=24_pool_shape=(2__2)_max=True_filter_shapes=[(5__5)__(3__3)]_dropout=0.5 \\
}\lmdkeysquarevgg


\pgfplotstableread[row sep=\\,col sep=&]{
Testset & Runs & Parameters & Acc1 & Std1 & Model \\
ballroom         & 5 &      19268 & 0.14183 & 0.02673 & shallow_key_in=(40__256__1)_out=256_filters=1_short_shape=(3__1)_long_shape=(40__1)_dropout=0.1 \\
ballroom         & 5 &      43400 & 0.13467 & 0.07634 & shallow_key_in=(40__256__1)_out=256_filters=2_short_shape=(3__1)_long_shape=(40__1)_dropout=0.3 \\
ballroom         & 5 &     107024 & 0.12350 & 0.04188 & shallow_key_in=(40__256__1)_out=256_filters=4_short_shape=(3__1)_long_shape=(40__1)_dropout=0.3 \\
ballroom         & 5 &     191128 & 0.17450 & 0.04136 & shallow_key_in=(40__256__1)_out=256_filters=6_short_shape=(3__1)_long_shape=(40__1)_dropout=0.3 \\
ballroom         & 5 &     295712 & 0.16304 & 0.06201 & shallow_key_in=(40__256__1)_out=256_filters=8_short_shape=(3__1)_long_shape=(40__1)_dropout=0.1 \\
ballroom         & 5 &     566320 & 0.16734 & 0.05722 & shallow_key_in=(40__256__1)_out=256_filters=12_short_shape=(3__1)_long_shape=(40__1)_dropout=0.5 \\
}\ballroomtempokeyshallow

\pgfplotstableread[row sep=\\,col sep=&]{
Testset & Runs & Parameters & Acc1 & Std1 & Model \\
ballroom         & 5 &      17988 & 0.13954 & 0.02338 & shallow_key_in=(40__256__1)_out=256_filters=1_short_shape=(3__1)_long_shape=(20__1)_dropout=0.3 \\
ballroom         & 5 &      38280 & 0.15903 & 0.04419 & shallow_key_in=(40__256__1)_out=256_filters=2_short_shape=(3__1)_long_shape=(20__1)_dropout=0.3 \\
ballroom         & 5 &      86544 & 0.15330 & 0.06078 & shallow_key_in=(40__256__1)_out=256_filters=4_short_shape=(3__1)_long_shape=(20__1)_dropout=0.3 \\
ballroom         & 5 &     145048 & 0.11805 & 0.02840 & shallow_key_in=(40__256__1)_out=256_filters=6_short_shape=(3__1)_long_shape=(20__1)_dropout=0.1 \\
ballroom         & 5 &     213792 & 0.17765 & 0.03218 & shallow_key_in=(40__256__1)_out=256_filters=8_short_shape=(3__1)_long_shape=(20__1)_dropout=0.5 \\
ballroom         & 5 &     382000 & 0.15387 & 0.07222 & shallow_key_in=(40__256__1)_out=256_filters=12_short_shape=(3__1)_long_shape=(20__1)_dropout=0.5 \\
}\ballroomtempokeyshallowhalf

\pgfplotstableread[row sep=\\,col sep=&]{
Testset & Runs & Parameters & Acc1 & Std1 & Model \\
ballroom         & 5 &      33092 & 0.62436 & 0.26416 & shallow_tempo_in=(40__256__1)_out=256_filters=1_short_shape=(1__3)_long_shape=(1__256)_dropout=0.1 \\
ballroom         & 5 &      98696 & 0.84928 & 0.03781 & shallow_tempo_in=(40__256__1)_out=256_filters=2_short_shape=(1__3)_long_shape=(1__256)_dropout=0.3 \\
ballroom         & 5 &     328208 & 0.86390 & 0.01322 & shallow_tempo_in=(40__256__1)_out=256_filters=4_short_shape=(1__3)_long_shape=(1__256)_dropout=0.5 \\
ballroom         & 5 &     688792 & 0.87564 & 0.00899 & shallow_tempo_in=(40__256__1)_out=256_filters=6_short_shape=(1__3)_long_shape=(1__256)_dropout=0.5 \\
ballroom         & 5 &    1180448 & 0.86934 & 0.01557 & shallow_tempo_in=(40__256__1)_out=256_filters=8_short_shape=(1__3)_long_shape=(1__256)_dropout=0.3 \\
ballroom         & 5 &    2556976 & 0.87851 & 0.02289 & shallow_tempo_in=(40__256__1)_out=256_filters=12_short_shape=(1__3)_long_shape=(1__256)_dropout=0.5 \\
}\ballroomtempotemposhallow

\pgfplotstableread[row sep=\\,col sep=&]{
Testset & Runs & Parameters & Acc1 & Std1 & Model \\
ballroom         & 5 &      24900 & 0.62579 & 0.27965 & shallow_tempo_in=(40__256__1)_out=256_filters=1_short_shape=(1__3)_long_shape=(1__128)_dropout=0.3 \\
ballroom         & 5 &      65928 & 0.82865 & 0.02674 & shallow_tempo_in=(40__256__1)_out=256_filters=2_short_shape=(1__3)_long_shape=(1__128)_dropout=0.5 \\
ballroom         & 5 &     197136 & 0.88252 & 0.02125 & shallow_tempo_in=(40__256__1)_out=256_filters=4_short_shape=(1__3)_long_shape=(1__128)_dropout=0.3 \\
ballroom         & 5 &     393880 & 0.87880 & 0.02487 & shallow_tempo_in=(40__256__1)_out=256_filters=6_short_shape=(1__3)_long_shape=(1__128)_dropout=0.5 \\
ballroom         & 5 &     656160 & 0.88395 & 0.01464 & shallow_tempo_in=(40__256__1)_out=256_filters=8_short_shape=(1__3)_long_shape=(1__128)_dropout=0.3 \\
ballroom         & 5 &    1377328 & 0.87077 & 0.01339 & shallow_tempo_in=(40__256__1)_out=256_filters=12_short_shape=(1__3)_long_shape=(1__128)_dropout=0.5 \\
}\ballroomtempotemposhallowhalf

\pgfplotstableread[row sep=\\,col sep=&]{
Testset & Runs & Parameters & Acc1 & Std1 & Model \\
ballroom         & 5 &       9322 & 0.72292 & 0.02996 & vgg_like_in=(40__256__1)_out=256_filters=2_pool_shape=(2__2)_max=True_filter_shapes=[(1__5)__(1__3)]_dropout=0.1 \\
ballroom         & 5 &      27228 & 0.78911 & 0.02974 & vgg_like_in=(40__256__1)_out=256_filters=4_pool_shape=(2__2)_max=True_filter_shapes=[(1__5)__(1__3)]_dropout=0.1 \\
ballroom         & 5 &      89560 & 0.83811 & 0.02205 & vgg_like_in=(40__256__1)_out=256_filters=8_pool_shape=(2__2)_max=True_filter_shapes=[(1__5)__(1__3)]_dropout=0.1 \\
ballroom         & 5 &     320304 & 0.88195 & 0.02429 & vgg_like_in=(40__256__1)_out=256_filters=16_pool_shape=(2__2)_max=True_filter_shapes=[(1__5)__(1__3)]_dropout=0.1 \\
ballroom         & 5 &     692488 & 0.87851 & 0.02327 & vgg_like_in=(40__256__1)_out=256_filters=24_pool_shape=(2__2)_max=True_filter_shapes=[(1__5)__(1__3)]_dropout=0.3 \\
}\ballroomtempotempovgg

\pgfplotstableread[row sep=\\,col sep=&]{
Testset & Runs & Parameters & Acc1 & Std1 & Model \\
ballroom         & 5 &       9322 & 0.25444 & 0.04887 & vgg_like_in=(40__256__1)_out=256_filters=2_pool_shape=(2__2)_max=True_filter_shapes=[(5__1)__(3__1)]_dropout=0.1 \\
ballroom         & 5 &      27228 & 0.34441 & 0.07973 & vgg_like_in=(40__256__1)_out=256_filters=4_pool_shape=(2__2)_max=True_filter_shapes=[(5__1)__(3__1)]_dropout=0.1 \\
ballroom         & 5 &      89560 & 0.50917 & 0.04661 & vgg_like_in=(40__256__1)_out=256_filters=8_pool_shape=(2__2)_max=True_filter_shapes=[(5__1)__(3__1)]_dropout=0.1 \\
ballroom         & 5 &     320304 & 0.52550 & 0.08624 & vgg_like_in=(40__256__1)_out=256_filters=16_pool_shape=(2__2)_max=True_filter_shapes=[(5__1)__(3__1)]_dropout=0.1 \\
ballroom         & 5 &     692488 & 0.59628 & 0.09078 & vgg_like_in=(40__256__1)_out=256_filters=24_pool_shape=(2__2)_max=True_filter_shapes=[(5__1)__(3__1)]_dropout=0.1 \\
}\ballroomtempokeyvgg

\pgfplotstableread[row sep=\\,col sep=&]{
Testset & Runs & Parameters & Acc1 & Std1 & Model \\
ballroom         & 5 &       7134 & 0.68481 & 0.01310 & vgg_like_in=(40__256__1)_out=256_filters=1_pool_shape=(2__2)_max=True_filter_shapes=[(5__5)__(3__3)]_dropout=0.1 \\
ballroom         & 5 &      23082 & 0.79398 & 0.02974 & vgg_like_in=(40__256__1)_out=256_filters=2_pool_shape=(2__2)_max=True_filter_shapes=[(5__5)__(3__3)]_dropout=0.1 \\
ballroom         & 5 &      82188 & 0.88940 & 0.02259 & vgg_like_in=(40__256__1)_out=256_filters=4_pool_shape=(2__2)_max=True_filter_shapes=[(5__5)__(3__3)]_dropout=0.1 \\
ballroom         & 5 &     309240 & 0.91777 & 0.01015 & vgg_like_in=(40__256__1)_out=256_filters=8_pool_shape=(2__2)_max=True_filter_shapes=[(5__5)__(3__3)]_dropout=0.3 \\
ballroom         & 5 &    1198704 & 0.92436 & 0.01679 & vgg_like_in=(40__256__1)_out=256_filters=16_pool_shape=(2__2)_max=True_filter_shapes=[(5__5)__(3__3)]_dropout=0.3 \\
ballroom         & 5 &    2668648 & 0.92607 & 0.01244 & vgg_like_in=(40__256__1)_out=256_filters=24_pool_shape=(2__2)_max=True_filter_shapes=[(5__5)__(3__3)]_dropout=0.5 \\
}\ballroomtemposquarevgg


\begin{table*}[t]
\centering
\subfloat[Tempo]{
\begin{tabular}{rcccc} \toprule
Architecture        & GS & GT & LMD & BR \\ \midrule
\shallowtempo       & 86.5 \tiny{1.5} & 60.3 \tiny{2.7} & 94.0 \tiny{1.0} & 87.9 \tiny{2.3} \\
\deeptempo          & \textbf{88.7} \tiny{0.6} & 63.1 \tiny{0.6} & 94.5 \tiny{0.7} & 88.2 \tiny{2.4}\\
\shallowkey         &  4.5 \tiny{1.9} & 11.5 \tiny{1.3} &  9.4 \tiny{2.1} & 16.7 \tiny{5.7} \\
\deepkey            & 49.6 \tiny{2.5} & 40.2 \tiny{1.4} & 73.0 \tiny{2.4} & 59.6 \tiny{9.1} \\
\deepsquare         & 88.1 \tiny{1.3} & \textbf{64.7} \tiny{2.1} & \textbf{96.2} \tiny{0.4} & \textbf{92.4} \tiny{1.7} \\ \midrule
Literature          & 82.5\cite{Schreiber2018a} & 78.3\cite{percival2014streamlined} & --- & 92.0\cite{Schreiber2018a} \\ 
\bottomrule
\end{tabular}
}
\quad
\subfloat[Key]{
\begin{tabular}{ccc} \toprule
GS & GT & LMD \\ \midrule
1.7 \tiny{0.4} &  4.9 \tiny{0.7} & 11.0 \tiny{3.7} \\
46.8 \tiny{4.3} & 38.4 \tiny{2.4} & 60.7 \tiny{0.4} \\
50.8 \tiny{3.8} & 43.8 \tiny{1.4} & 67.1 \tiny{0.9} \\
55.4 \tiny{2.7} & 44.8 \tiny{2.0} & \textbf{71.3} \tiny{0.2} \\
\textbf{58.5} \tiny{3.9} & \textbf{49.9} \tiny{2.0} & 68.9 \tiny{2.5} \\ \midrule
67.9\cite{Korzeniowski2018} & \texttildelow45\cite{Kraft2013tonalness} & --- \\
\bottomrule
\end{tabular}
}
\caption{Mean estimation accuracies of 5 runs with standard deviation (small font).
Best results per test are set in \textbf{bold}.
Model variants chosen based on validation performance (ignoring parameter count).
GS=\giantsteps, GT=\gtzan, LMD=\lmd, BR=\ballroom.}
\label{tab:test}
\end{table*}

\section{Discussion}
\label{sec:discussion}

The results show that simple shallow networks with axis-aligned, directional filters
can perform well on both the key and tempo task. Conceptually, both tasks are similar enough
that virtually the same architecture can be used for either one, as long as the input
representation and the filter direction are appropriate.
Using the wrong filter direction, e.g., \shallowkey\ for the tempo task, leads to
very poor results close to the random baseline.
Together, this strongly supports the hypothesis that the \shallow\
architecture indeed learns what we want it to learn, i.e., pitch patterns for key detection
or rhythmic patterns for tempo detection, but not both.

This stands in contrast to the standard VGG-style network~(\deepsquare). Because
of its square filters, we cannot be certain what it learns, just by analyzing
its static architecture. It is designed to pick up on anything that could provide a hint
towards correct classification, be it rhythm and pitch patterns, or timbral
properties like instrumentation. And indeed our experiment shows that without being specialized
for either key or tempo estimation in any way, \deepsquare\ works very well for both tasks.
In \secref{sec:results} we noted that \deepsquare\ achieved the greatest tempo accuracy
for \ballroom\ and the greatest key accuracy for \gtzankey\ by a considerable margin
of $4.2\,\mathrm{pp}$ and $5.1\,\mathrm{pp}$, respectively.
This margin may be a result of the fact that key and tempo are
related to genre~\cite{horschlager2015,SchullerER08_tangoWaltz_EURASIP,Xiao2008,Faraldo2016}.
Specifically, in \ballroom\ there is a strong correlation between genre and tempo,
and \gtzankey\ is the key test set with the greatest genre diversity and therefore
stands to benefit the most from an architecture that can distinguish genres based on
\emph{both} temporal and timbral properties.
Consequently, square filters \emph{should} improve accuracy results for these datasets.
But this does not conclusively show that only the network's ability to measure
specifically key or tempo is reflected by these results, as the
system is by design vulnerable to confounds~\cite{Sturm2014simple}.
By using directional filters in \deepkey\ and \deeptempo\ we intentionally limit the
standard VGG-style architecture in a way that seeks to lessen this vulnerability
as well as reduce the number of required parameters.

The results for \deepkey\ and \deeptempo\ show that a VGG-style network with directional
filters can perform very well on either task. For networks with a large number of
parameters test results are similar to \deepsquare, with a tendency towards a slightly
worse performance. Interestingly, the situation is different for
low-capacity networks with $k=2$ for \deepkey, and $k=1$ for \deepsquare.
Here, \deepkey\ clearly outperforms \deepsquare, even though the
parameter count is similar. Perhaps with ca.\ $5\,000$ parameters
\deepsquare\ simply does not have enough capacity aligned
in the right direction to still perform well on the task.

The fact that \deepkey\ and \deeptempo\ with $k=2$ perform very poorly on the tasks
they are not meant for, supports the hypothesis that they only learn the intended
features for the tasks they are meant for. For $k>2$ we cannot be quite as certain,
as both architectures reach higher accuracy scores on the tasks they were not meant
for for greater values of $k$. We believe this effect may be a result of the $2\times2$
max pooling in the \deepmod\ modules.

\section{Conclusions}
\label{sec:conclusions}

We have shown that shallow, signal processing-inspired CNN
architectures using directional filters can be used successfully for both tempo and
key detection. 
By using shallow networks designed for key detection on the tempo task and vice versa, we were
able to experimentally support the hypothesis that these networks are incapable of matching
information from the domain they were not meant for, which would make them less susceptible to
confounds.

We further demonstrated that a standard VGG-style architecture can be used for
tempo estimation, as it has been shown before for key detection~\cite{Korzeniowski2018}.
By replacing square filters with directional filters, we derived a musically motivated,
directional VGG-variant that performs similarly well as the original one, but
is less vulnerable to confounds, especially when used for key detection with
low capacity models. In such scenarios we were also able to observe efficiency gains,
i.e., better performance than the standard VGG-style network with similar parameter
counts.

\begin{additional}
Code to recreate models and reproduce the reported results can be found at
\url{https://github.com/hendriks73/directional_cnns}.
\end{additional}

\begin{acknowledgments}
The International Audio Laboratories Erlangen are a joint institution of
the Friedrich-Alexander-Universit\"at Erlangen-N\"urnberg (FAU) and
Fraunhofer Institute for Integrated Circuits IIS. Meinard M\"uller is
supported by the German Research Foundation (DFG MU 2686/11-1).
\end{acknowledgments} 

\bibliography{referencesMusic}

\end{document}